\pdfoutput=1
\documentclass[runningheads]{llncs}

\usepackage{versions}
	\includeversion{extended}

\usepackage[dvipsnames]{xcolor}
\usepackage[utf8]{inputenc}

\usepackage{microtype}
\usepackage{latexsym}
	\usepackage{amssymb}
	\usepackage{bm} 
\usepackage{mathtools} 
\usepackage[inline,shortlabels]{enumitem} 
	\setlist[itemize]{topsep=0.2ex}
\usepackage{booktabs} 
\usepackage{tikz} 
	\usetikzlibrary{automata,positioning,shapes,fit}
	\usetikzlibrary{calc}
	\usetikzlibrary{svg.path}
	\usetikzlibrary{captl} 
	\usetikzlibrary{extshapes}
\usepackage{pgfplots} 
	\pgfplotsset{compat=1.12}	
	\usepgfplotslibrary{units}  
	\usepgfplotslibrary{fillbetween} 
	\usetikzlibrary{patterns} 
\usepackage{xargs}
\usepackage{wrapfig}
\usepackage{cutwin}
\usepackage[ruled,vlined,linesnumbered]{myalgorithm2e}
\usepackage{graphicx}
	\graphicspath{ {figs/} }
\usepackage{etoolbox}
\usepackage{upgreek}
\usepackage{pifont} 
\usepackage{relsize} 
\usepackage{fancyhdr} 
	\usepackage[yyyymmdd,hhmmss]{datetime}
	
\usepackage{soul} 
\usepackage[tight,footnotesize]{subfigure}
\usepackage{stackengine}
\usepackage{printlen}
\usepackage[all,cmtip]{xy}
\usepackage[ligature, inference]{semantic}
\usepackage{latexsym}
\usepackage{courier} 
\usepackage{scalerel}
\usepackage{inconsolata} 

\newcommand{\HL}[1]{{ #1 }}

\newcommand{\tabref}		[1]{Table~\ref{#1}}
\newcommand{\figref}		[1]{Fig.~\ref{#1}}
\newcommand{\thmref}		[1]{Theorem~\ref{#1}}

\newcommand{\lemref}		[1]{Lemma~\ref{#1}}
\newcommand{\exref}			[1]{Example~\ref{#1}}
\newcommand{\algref}		[1]{Algorithm~\ref{#1}}
\newcommand{\secref}		[1]{Sec.~\ref{#1}}

\newcommand{\defref}		[1]{Definition~\ref{#1}}

\newcommand{\ie}{i.e., }
\newcommand{\eg}{e.g., }
\newcommand{\suchthat}{\mbox{s.t.}}

\newcommand{\Eff}[1]{%
		\ifthenelse{\isempty{#1}}%
		{\mathit{Eff}}
		{\mathit{Eff}\!\!\left(#1\right)}}

\newcommand {\Last}{\mathit{last}}

\newcommand {\Probability}{\mathrm{Pr}}
\newcommand {\Prob}{\mathrm{Pr}}
\newcommand {\Pquery}{\mathbb{P}}

\newcommand {\Naturals}{\mathbb{N}}
\newcommand	{\Reals}	{\mathbb{R}}
\newcommand {\Booleans}{\mathbb{B}}

\newcommand	{\PRISMG}	{{PRISM-games}}
\newcommand {\UPPAALS}{{\textsc{Uppaal Stratego}}}
\newcommand {\MATLAB}{{MATLAB}}

\newcommand {\Always}{\square}
\newcommand {\Eventually}{\lozenge}
\newcommand {\Until}{\mathsf{U}}
\newcommand {\Weakuntil}{\mathsf{W}}
\newcommand {\Next}{\mathsf{X}}

\newcommand {\Eq}{\!=\!}

\newcommand{\Ipath}{\mathit{IPath}}
\newcommand{\Fpath}{\mathit{FPath}}
\newcommand {\Reach}{\mathit{Reach}}

\newcommand{\Act}{\mathit{Act}}
\newcommand{\AP}{\mathit{AP}}
\newcommand{\Paths}{\mathit{Paths}}
\newcommand{\Pmatrix}{\mathbf{P}}
\newcommand{\TaskSet}{\Xi} 

\newcommand{\Pset}{\mathcal{P}}

\newcommand{\Protocol}{\Pi}
\newcommand{\Protocolset}{\mathfrak{P}}
\newcommand{\Str}{\sigma}
\newcommand{\Strset}{\Sigma}
\newcommand{\Path}{\pi}
\newcommand{\Opt}{\mathrm{opt}}
\def\opt{\operatornamewithlimits{opt}}

\newcommand{\assign}{\leftarrow} 

\newcommand{\LGoal}{\mathit{goal}}
\newcommand{\LChrg}{\mathit{chrg}}
\newcommand{\LOn}{\mathit{on}}
\newcommand{\LErr}{\mathit{error}}
\newcommand{\LSleep}{\mathit{sleep}}
\newcommand{\LSafe}{\mathit{safe}}

\newcommand{\ANorth}{\mathsf{N}}
\newcommand{\ASouth}{\mathsf{S}}
\newcommand{\AEast}{\mathsf{E}}
\newcommand{\AWest}{\mathsf{W}}
\newcommand{\ASleep}{\mathsf{sleep}}
\newcommand{\AError}{\mathsf{error}}

\newcommand{\ANorthSym}{{\color{RoyalBlue}\blacktriangle}}

\newcommand{\AEastSym} {{\color{RoyalBlue}\blacktriangleright}}

\newcommand{\AWab}{{\color{RedOrange}\mathbin{\rotatebox[origin=c]{180}{$\triangledown$}}}}
\newcommand{\AWac}{{\color{RedOrange}\triangledown}}
\newcommand{\AWbd}{{\color{RedOrange}\triangleright}}

\newcommand{\Domain}{\mathit{dom}}
\newcommand{\defined}{\downarrow}
\newcommand{\undefined}{\uparrow}

\newcommand{\VRepeat}{\mathit{repeat}}

\newcommand{\Synth}{\textsc{Synth}}
\newcommand{\Reachset}{\textsc{Reach}}
\newcommand{\ReachPset}{\textsc{ReachP}}
\newcommand{\Verify}{\textsc{Verify}}

\newcommand{\Post}{\mathit{Post}}

\renewcommand {\iff}{\Leftrightarrow}

\newcommand{\Plusplus}{{\raisebox{.2\height}{\scalebox{.8}{++}}}}

\makeatletter
\newsavebox\myboxA
\newsavebox\myboxB
\newlength\mylenA
\newcommand*\xoverline[2][0.75]{%
    \sbox{\myboxA}{$\m@th#2$}%
    \setbox\myboxB\null
    \ht\myboxB=\ht\myboxA%
    \dp\myboxB=\dp\myboxA%
    \wd\myboxB=#1\wd\myboxA
    \sbox\myboxB{$\m@th\overline{\copy\myboxB}$}
    \setlength\mylenA{\the\wd\myboxA}
    \addtolength\mylenA{-\the\wd\myboxB}%
    \ifdim\wd\myboxB<\wd\myboxA%
       \rlap{\hskip 0.5\mylenA\usebox\myboxB}{\usebox\myboxA}%
    \else
        \hskip -0.5\mylenA\rlap{\usebox\myboxA}{\hskip 0.5\mylenA\usebox\myboxB}%
    \fi}
\makeatother

\newcommand{\ActW}{\xoverline[0.8]{\Act}}
\newcommand{\Playerset}{\Gamma}
\newcommand{\Trace}{\mathit{trace}}

\newcommand{\St}[1]{{\langle #1 \rangle}}
\newcommand{\ra}{\rightarrow}
\newcommand{\eqs}{\!\!=\!\!}
\let\xra\xrightarrow
\let\var\mathsf

\newcommand*\circled[1]{{\tikz[baseline=(char.base)]{%
	\node[shape=circle,draw,inner sep=0.5pt] (char) {\footnotesize\textup{#1}};}}}
\newcommand*{\Pone}{\circled{1}}
\newcommand*{\Ptwo}{\circled{2}}

\newcommand{\MDP}{\mathcal{M}}
\newcommand{\CAPTL}{\mathcal{A}}
\newcommand{\SRS}{{\MDP_{\CAPTL}^{\Str}}}
\newcommand{\MDPxPRO}{{\MDP^{\Protocol}}}
\newcommand{\hpi}{\hat{\pi}}
\newcommand{\qact}{{q_{\mathrm{act}}}}

\newcommand{\MDPx}{\hat{\MDP}}
\newcommand{\CAPTLx}{\hat{\CAPTL}}
\newcommand{\Strx}{\hat{\Str}}

\newcommand{\MDPdef}{%
	( S, \allowbreak \Act, \allowbreak \Pmatrix, \allowbreak s_0,
	\allowbreak \AP, \allowbreak L )%
	}
\newcommand{\CAPTLdef}{%
	( Q, \allowbreak W, \allowbreak \TaskSet, \allowbreak \hookrightarrow,
	\allowbreak q_0 )%
	}

\newcommand\restr[2]{{
  \left.\kern-\nulldelimiterspace 
  #1 
  \vphantom{\big|} 
  \right|_{#2} 
  }}

	\let\emptyset\varnothing
\newlength{\xywd}

\renewcommand\ldots{\makebox[1em][c]{.\hfil.\hfil.}}

\renewcommand{\vec}[1]{\mathbf{#1}}

\SetKwComment{Comment}{$\triangleright$\ }{}

\SetSideCommentRight
\SetCommentSty{text}
\SetKw{Break}{break}
\SetKwFor{Construct}{construct}{such that}{}{}
\SetKwRepeat{Do}{do}{while}
\DontPrintSemicolon

\newcommand{\RemoveAlgExtraSpace}{\vspace{-0.6ex}}

\setpremisesend{0.50em} 
\newcommand{\Rule}[1]{{$\mathrm{[R#1]}$}}

\let\kw\texttt

\tikzset{pAction/.style={font={\fontsize{7}{0}\selectfont}, RoyalBlue, above, pos=.5}}
\tikzset{pGuard/.style={font={\fontsize{7}{0}\selectfont}, Red}}
\tikzset{pLabel/.style={font={\fontsize{7}{0}\selectfont}, Black!90}}
\tikzset{pDistr/.style={font={\fontsize{7}{0}\selectfont}, Black!90}}
\tikzset{pAssign/.style={font={\fontsize{7}{0}\selectfont}, ForestGreen}}
\tikzset{pArc/.style={-,shorten <=0pt,shorten >=0pt}}

\newcommand{\RemoveSpaceAfterTikz}{\vspace{-6pt}}
\newcommand{\kwm}[1]{\mathit{#1}}
\newcommand{\kwt}[1]{\textit{#1}}

\newcount\ppnum
\newcommand\num[1]{%
        \ppnum=#1\relax
        \ifnum\ppnum<0
                $-$%
                \ppnum=-\ppnum
        \fi
        \let\pptemp\empty
        \loop\ifnum\ppnum>999
                \count255=\ppnum
                \divide\ppnum by1000
                \count255=\numexpr \count255 - 1000*\ppnum \relax
                \edef\pptemp{,\ifnum\count255<100 0\ifnum\count255<10 0\fi\fi
                             \the\count255 \pptemp}%
        \repeat
        \the\ppnum
        \pptemp
}

\definecolor{orcidlogocol}{HTML}{A6CE39}
\tikzset{
  orcidlogo/.pic={
    \fill[orcidlogocol] svg{M256,128c0,70.7-57.3,128-128,128C57.3,256,0,198.7,0,128C0,57.3,57.3,0,128,0C198.7,0,256,57.3,256,128z};
    \fill[white] svg{M86.3,186.2H70.9V79.1h15.4v48.4V186.2z}
                 svg{M108.9,79.1h41.6c39.6,0,57,28.3,57,53.6c0,27.5-21.5,53.6-56.8,53.6h-41.8V79.1z M124.3,172.4h24.5c34.9,0,42.9-26.5,42.9-39.7c0-21.5-13.7-39.7-43.7-39.7h-23.7V172.4z}
                 svg{M88.7,56.8c0,5.5-4.5,10.1-10.1,10.1c-5.6,0-10.1-4.6-10.1-10.1c0-5.6,4.5-10.1,10.1-10.1C84.2,46.7,88.7,51.3,88.7,56.8z};
  }
}

\newcommand\orcidicon[1]{\href{https://orcid.org/#1}{\mbox{\scalerel*{
\begin{tikzpicture}[yscale=-1,transform shape]
\pic{orcidlogo};
\end{tikzpicture}
}{|}}}}

\spnewtheorem*{proofsketch}{Proof Sketch}{\itshape}{\rmfamily}
\let\qedsymbol\qed
\AtEndEnvironment{proof}{\null\hfill\qedsymbol}
\AtEndEnvironment{proofsketch}{\null\hfill\qedsymbol}

\begin{document}

\title{Context-Aware Temporal Logic\\
 	for Probabilistic Systems%
 	}
%
%

\author{%
	Mahmoud Elfar%
		\orcidID{0000-0002-5579-1255}%
	\and
	Yu Wang%
		\orcidID{0000-0002-0431-1039}%
	\and
	Miroslav Pajic%
		\orcidID{0000-0002-5357-0117}%
	}
\authorrunning{M. Elfar et al.}
%
\institute{%
	Duke University, Durham NC 27708, USA\\
	\email{\{mahmoud.elfar,yu.wang94,miroslav.pajic\}@duke.edu}\\
	\url{http://cpsl.pratt.duke.edu}%
}
%
%
\maketitle              

\begin{abstract}

In this paper, we \HL{introduce} the context-aware probabilistic temporal logic (CAPTL) that provides an intuitive way to formalize system requirements by a set of PCTL objectives with a context-based priority structure. We formally present the syntax and semantics of CAPTL and propose a synthesis algorithm for CAPTL requirements. We also implement the algorithm based on the PRISM-games model checker. Finally, we demonstrate the usage of CAPTL on two case studies: a robotic task planning problem, and synthesizing error-resilient scheduler for micro-electrode-dot-array digital microfluidic biochips.

	
\end{abstract}


\renewcommand{\xrightarrow}[2][]{%
  \sbox{0}{$\scriptstyle#2$}%
  \xywd=\wd0%
  \!\!\!\xymatrix@C\dimexpr\xywd+0.6em\relax{{}\ar[r]^{\!\!#2}&{\!\!\!}\!}\!\!%
}




\makeatletter
\setlength\textfloatsep{6mm\@plus 2\p@ \@minus 4\p@}
\makeatother




\setlength{\abovedisplayskip}{0.4ex plus 0.2ex minus 0.1ex}
\setlength{\belowdisplayskip}{0.4ex plus 0.2ex minus 0.1ex}
\makeatletter
\renewcommand\subsubsection{\@startsection{subsubsection}{3}{\z@}%
                       {-8\p@ \@plus -4\p@ \@minus -4\p@}%
                       {-0.5em \@plus -0.22em \@minus -0.1em}%
                       {\normalfont\normalsize\bfseries\boldmath}}
\renewcommand\section{\@startsection{section}{1}{\z@}%
                       {-10\p@ \@plus -4\p@ \@minus -4\p@}%
                       {8\p@ \@plus 4\p@ \@minus 6\p@}%
                       {\normalfont\large\bfseries\boldmath
                        \rightskip=\z@ \@plus 8em\pretolerance=10000 }}
\makeatother

\setlength{\topsep}{4pt plus 4pt minus 6pt}






\section{Introduction}
\label{sec:intro}

The correct-by-design paradigm in Cyber-Physical Systems (CPS) has been a central concept
 during the design phase of various system components.
This paradigm requires the abstraction of both the system behavior and the design
 requirements~\HL{\cite{neema2003constraint,pajic2012model}}.
Typically, the system behavior is modeled as a discrete Kripke structure, with nondeterministic
 transitions representing various actions or choices that need to be resolved.
In systems where probabilistic behavior is prevalent, formalisms such as Markov decision
 processes (MDPs) are best suited.
The applications of correct-by-design synthesis paradigm span CPS fields such as
 robot path and behavior planning~\HL{\cite{bozkurt2020control,kress2018synthesis},}
 smart power grids~\cite{puggelli2014robust},
 safety-critical medical devices~\cite{jiang2012modeling}, and
 autonomous vehicles~\cite{seshia2015formal}.

Temporal logic (TL) can be utilized to formalize
 CPS design requirements.
For example, Linear Temporal Logic \HL{(LTL)~\cite{baier2008principles}}
 is used to capture safety and reachability
 requirements over Boolean predicates defined over the state space.
Similarly, computation tree logic \HL{(CTL)~\cite{baier2008principles}}
 allows for expressing requirements 
 over all computations branching from a given state.
Probabilistic computation tree logic (PCTL) can be viewed as
a probabilistic variation of CTL to  
reason about the satisfaction probabilities of temporal requirements.

The choice of which TL to use 
 is both a science and an art.
Nevertheless, fundamental factors include expressiveness
 (\ie whether the design requirements of interest can be expressed by the logic),
 and the existence of model checkers that can verify the system model against
 the design requirement, synthesize winning strategies, or generate
 counterexamples.
Although prevalent TLs can be inherently expressive, two notions
 are oftentimes overlooked, namely, how easy it is to correctly formalize
 the design requirements,
 and whether existing model checkers are optimized for such requirements.
The more complex it becomes to formalize a given requirement, the more likely it is
 that human error is introduced in the process.

In particular, we focus in this paper on requirements that are naturally specified as 
 a set of various objectives with an underlying priority structure.
For instance, the objective of an embedded controller might be focused
 on achieving a primary task. However, whenever the chances of achieving such task 
 fall below a certain threshold, the controller shall proceed with 
 a fail-safe procedure.
Such requirement, while being easy to state and understand, can prove challenging
 when formalized for two reasons.
First, multiple objectives \HL{might be} involved with a priority structure,
 \ie one objective takes priority over another.
Second, the context upon which the objectives are switched is of probabilistic
 nature, \ie it requires the ability to prioritize objectives based on probabilistic
 invariants.

To this end, in this work we consider the problem of modeling and synthesis of CPS
 modeled as MDPs,
 with context-based probabilistic requirements,
 where a context is defined over probabilistic conditions.
We tackle this problem by introducing the context-aware probabilistic temporal
 logic (CAPTL).
CAPTL provides intuitive means to formalize design requirements as a set of objectives
 with a priority structure.
For example, a requirement can be defined in terms of primary and secondary objectives,
 where switching from the former to the latter is based upon a probabilistic condition
 (\ie a context).
The ability to define context as probabilistic conditions sets CAPTL apart from similar
 TLs.

In addition to providing the syntax and semantics of CAPTL for MDPs, we investigate
 the problem of synthesizing winning strategies based on CAPTL requirements.
Next, we demonstrate how the synthesis problem can be reduced to 
 a set of PCTL-based synthesis sub-problems.
Moreover, for deterministic CAPTL requirements with persistence objectives, we propose
 an optimized synthesis algorithm.
Finally, we implement the algorithm on top of \PRISMG{}~\cite{kwiatkowska2018prism}, 
 and we show experimental results for two case studies where we synthesize a robotic
 task planner, and an error-resilient scheduler for microfluidic biochips.

\subsubsection{Organization.}
The rest of this section discusses related work.
Preliminaries and a motivating example are provided in~\secref{sec:problem}.
In~\secref{sec:logic} we introduce the syntax and semantics of CAPTL.
The CAPTL-based synthesis problem is introduced in~\secref{sec:synthesis},
 where we first explore how a CAPTL requirement can be approached using
 PCTL, followed by our proposed synthesis algorithm.
For evaluation, we \HL{consider} two case studies in~\secref{sec:case}.
Finally, we conclude the paper in~\secref{sec:conclusion}.

\subsubsection{Related Work.}

The problem of multi-objective model checking and synthesis has been studied in literature,
 spanning both MDPs and stochastic games, for various properties, 
 including reachability, safety,
 probabilistic queries, and reward-based requirements%
 ~\cite{brenguier2016decidability,%
 etessami2007multi,%
 forejt2011quantitative,%
 forejt2012pareto}.
Our work improves upon the multi-objective synthesis paradigm
by enabling priorities over the multiple objectives as we will show in~\secref{sec:problem}.
One prevalent workaround is to define multiple reward structures, where states are assigned
 tuples of real numbers depicting how favorable they are with respect to multiple criteria.
The synthesis problem is then reduced to an optimization problem over either 
 a normalized version of the rewards (\ie assigning weights),
 or one reward with logical constraints on the
 others~\cite{baier2017synthesis,brazdil2016optimizing}.
Results are typically presented as Pareto curves, depicting feasible points in the
 reward space~\cite{forejt2012pareto}.
Our work differs in two aspects.
First, we use probabilities as means to define priorities rather
 than reward structures.
Second, the mechanics needed to define context-based priorities are an integral
 part of CAPTL.

Perhaps the closest notion to our context-based prioritization scheme 
 are probabilistic invariant sets (PIS)~\cite{kofman2012probabilistic}.
Both CAPTL and PIS involve the identification of state-space subsets that maintain
 a probability measure within specific bounds.
While prevalent in the field of probabilistic programs~\cite{barthe2016synthesizing},
 PIS was not considered in the field of CPS synthesis, despite the fact that
 (non-probabilistic) invariant sets are used in controller design~\cite{blanchini1999set}. 
The problem of merging strategies for MDPs that correspond to
 different objectives has been 
 investigated~\cite{boutilier1997prioritized,wiltsche2015assume}.
Our approach, however, is primarily focused on formalizing the notion of context-based
 priorities within the specification logic itself rather than altering the original model. 
While one can argue that PCTL alone can be used to define priorities by utilizing
 nested probabilistic operators, the nesting is typically limited to qualitative 
 operators~\cite{lahijanian2011control}.
In contrast, CAPTL relaxes such limitation by allowing quantitative operators as well.
Moreover, CAPTL-based synthesis provides an insight into which objective is being pursued
 at a given state.

\section{Problem Setting}
\label{sec:problem}

\subsubsection{Preliminaries.}

For a measurable event $E$, we denote its probability by $\Pr(E)$.
The powerset of $A$ is denoted by $\Pset(A)$. 
We use $\Reals$ and $\Booleans$ for the set of reals and booleans, respectively.
For a sequence or a vector $\pi$, we write $\pi[i]$, $i > 0$, to denote the 
 \HL{$i$-th} 
 element of $\pi$. 

We formally model the system as \HL{an MDP}.
MDPs feature both probabilistic and nondeterministic transitions,
 capturing both uncertain behaviors and nondeterministic choices in the modeled
 system, respectively.
We adopt the following definition for a system model as an MDP~\cite{baier2008principles}.
\begin{definition}[System Model] 
A system model is an MDP
 $\MDP = \MDPdef $ where
	$S$ is a finite set of states;
	$\Act $ is a finite set of actions;
	$\Pmatrix : S \times \Act \times S \rightarrow [0,1] $ is a transition
	probability function s.t. 
	\HL{$\sum_{s^\prime \in S} \Pmatrix (s,a,s^\prime) \in 
	\lbrace 0,1 \rbrace $ for $ a \in \Act $;} 
	$s_0$ is an initial state;
	$\AP$ is a set of atomic propositions; and
	$L : S \rightarrow \Pset(\AP)$ is a labeling function.
\end{definition}

Given a system $\MDP$,
a \emph{path} is a sequence of states $\pi = s_0 s_1 \dots $, such that
 $\Pmatrix (s_i, a_i, s_{i+1} ) > 0 $ where $a_i \in \Act(s_i)$ for all $i \geq 0$.
The trace of $\pi$ is defined as \HL{$\Trace(\pi) = L(s_0) L(s_1) \cdots $.}
We use $\Fpath_{\MDP,s} $ ($\Ipath_{\MDP,s}$) to denote the set of all finite
 (infinite) paths of $\MDP$ starting from $s \in S$.
We use $\Paths_{\MDP, s} $ to denote the set of \HL{all finite} and
 infinite paths starting from $s \in S$.
If $\Pmatrix (s,a,s^\prime) = p$ and $ p>0$,
 we write 
 {\larger[-1.5] $ s \xrightarrow{a,p} s^\prime $}
 to denote that, with probability $p$,
 taking action $ a $ in state $ s $ will yield to state $ s^\prime $.
We define the \emph{cardinality} of $\MDP$ as $| \MDP | = |S| + |\Pmatrix|$,
 where $|\Pmatrix|$ is the number of non-zero entries in $\Pmatrix$.

A \emph{strategy} (also known as a policy or a scheduler)
 defines the behavior upon which nondeterministic transitions in $\MDP$ are resolved.
A \emph{memoryless} strategy uses only the current state to determine what action to take,
 while a \emph{memory-based} strategy uses previous states as well.
We focus in this work on pure memoryless strategies, which 
 are \HL{shown} to suffice for PCTL reachability properties~\cite{baier2008principles}.
\begin{definition}[Strategy] 
A (pure memoryless) \emph{strategy} of 
 $\MDP = \MDPdef $
 is a function 
 $\Str : S \rightarrow \Act $
 that maps states to actions.
\end{definition}

By composing $\MDP$ and $\Str$, nondeterministic choices in $\MDP$ are resolved,
 reducing the model to a \emph{discrete-time Markov chain} (DTMC),
 denoted by $\MDP^\Str $. 
We use $\Prob_{\MDP,s}^{\Str}$ to denote the probability measure defined over
 the set of infinite paths $\Ipath_{\MDP,s}^\Str$.
The function $\Reach(\MDP,s,\Str)$ denotes the set of reachable states in $\MDP$
 starting from $s \in S$ under strategy $\Str$, while
 $\Reach(\MDP,s)$ denotes the set of all reachable states from $s$ under any strategy.

We use \emph{probabilistic computation tree logic} (PCTL) to formalize system objectives
 as temporal properties with probabilistic bounds, following the grammar
\begin{align*}
\Phi \Coloneqq 
	\top \mid a \mid \neg\Phi \mid \Phi\wedge\Phi \mid \Pquery_J [\varphi],
\quad
\varphi \Coloneqq 
	\Next \Phi \mid \Phi\Until\Phi \mid \Phi\Until^{\leq k}\Phi,
\end{align*}
where $J \subseteq [0,1]$, and $\Next$ and $\Until$ denote the \emph{next} and
 \emph{until} temporal modalities, respectively.
Other derived modalities include
 $\Eventually$ (\emph{eventually}),
 $\Always$ (\emph{always}), and
 $\Weakuntil$ (\emph{weak until}).
Given a system $\MDP$ and a strategy $\Str$, 
 \HL{the PCTL satisfaction semantics over $s\in S$ and
 $\pi \in \Paths_{\MDP,s}^{\Str}$ is defined
 as follows~\cite{baier2008principles,forejt2011automated}:} 
\begin{align*}
\begin{array}[t]{lllll}
		s,\Str & \models & a & \iff & a \in L(s) \\
		s,\Str & \models & \lnot \Phi & \iff & s \not\models \Phi \\
		s,\Str & \models & \Phi_1\!\land\!\Phi_2 & \iff & 
			s \models \Phi_1 \land s \models \Phi_2 \\
		s,\Str & \models & \Pquery_J \left[\varphi \right] & \iff & 
			\Pr \left\lbrace \Path \mid \Path \models \varphi \right\rbrace 
			\!\in\! J \\
		\Path,\Str & \models & \Next \Phi & \iff & \Path[1] \models \Phi \\
		\Path,\Str & \models & \Phi_1 \Until \Phi_2 & \iff &
			\HL{\exists j \!\in\! \Naturals, j \geq 0 .\, \Path[j] \models \Phi_2 \,}
			\land
			\HL{\left(\Path[k] \models \Phi_1, \forall k\in\Naturals, 0 \leq k < j \right)}
			\\
		\Path,\Str & \models & \Phi_1 \Until^{\leq n} \Phi_2 & \iff &
			\HL{\exists j \!\in\! \Naturals, 0\!\leq\! j\!\leq\! n .\, \Path[j] \models \Phi_2 \,}
			\land 
			\HL{\left(\Path[k] \models \Phi_1, \forall k\in\Naturals, 0 \!\leq\! k \!<\! j \right)}
\end{array}
\end{align*}

PCTL can be extended with \emph{quantitative queries} of the form 
 $\Pquery_{\min} [\varphi]$ ($\Pquery_{\max} [\Phi]$) to compute the minimum (maximum)
 probability of achieving $\varphi$%
 ~\cite{forejt2011automated,svorevnova2016quantitative},
 \ie
\begin{align*}
\Pquery_{\min}[\varphi] = \inf_{\Str \in \Strset} \Probability_{\MDP,s}^{\Str} 
	\left(\left\{ \pi \ | \ \pi \models \varphi \right\} \right) ,\;
\Pquery_{\max}[\varphi] = \sup_{\Str \in \Strset} \Probability_{\MDP,s}^{\Str} 
	\left(\left\{ \pi \ | \ \pi \models \varphi \right\} \right).
\end{align*}
We will denote such queries as $\Pquery_{\Opt}$ (read: optimal),
 where $\Opt \in \{ \max , \min \}$.

\subsubsection{Motivating Example.}

Consider the simple grid-world shown in~\figref{fig:example:a}.
The robot can move between rooms through doorways
 where obstacles can be probabilistically encountered (\eg closed doors),
 requiring the robot to consume more power.
The robot state is captured as a tuple $s: (\var{g},\var{h},\var{x},\var{y})$, where
 $\var{g} \in \{ \LOn, \LSleep, \LErr \} $ is the robot's status,
 $\var{h} \in \{0, 1, \ldots, 10 \}$ is the robot's battery level, and 
 $\var{x}$ and $\var{y}$
 are its current coordinates.
As shown in~\figref{fig:example:b}, 
 the system can be modeled as  
 $\MDP = \MDPdef $,
 where $\Act = \{ \ANorth, \ASouth, \AEast, \AWest, \ASleep, \AError \}$, and
 $s_0 = (0,10,1,1) $.
Suppose that the main objective for the robot is to reach the goal
 with a charge $h > 3$ (objective A).
However, if the probability of achieving objective A is less than $0.8$,
 the robot should prioritize reaching the charging station and switch to
 $\mathit{sleep}$ mode (objective B).
Moreover, if the probability of achieving objective B falls below $0.7$,
 the robot should stop and switch to $\mathit{err}$ mode, preferably in one of
 the safe zones (objective C).

\newcommand{\dashedactions}[1]{%
	\path[dashed] (#1) edge +(0:2em) node (tmp) {};
	\path[dashed] (#1) edge +(30:2em) node (tmp) {};
	\path[dashed] (#1) edge +(-30:2em) node (tmp) {};
	}

\begin{figure}[!t] 
	\centering
	\renewcommand{\thesubfigure}{(left)}%
	\subfigure{\includegraphics[width=0.35\columnwidth]{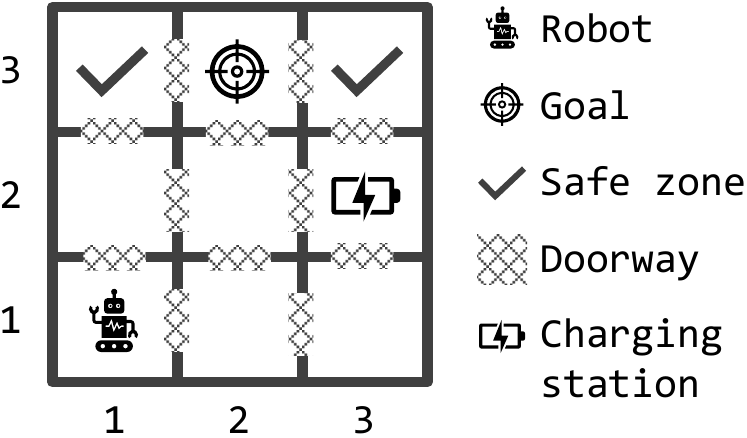}%
		\label{fig:example:a}}\hfill%
	\renewcommand{\thesubfigure}{(right)}%
	\subfigure{%
	\resizebox{0.64\columnwidth}{!}{\begin{tikzpicture}[automaton]
	\def\DistX{5.6}
	\def\DistY{1.7}
	\def\DeltaEq{0.6}
	\begin{scope}[shift={(0,0)}]
		\node[anchor=center] (legend) at (0,1) 
			{$s\colon \St{\var{g},\var{h},\var{x},\var{y}}$};
		\node[cstate,initial] (s0011) at (0,0) {$\St{0,10,1,1}$};
			\node[dot] (s0011e) at (1.5,0.6) {};
			\node[dot] (s0011n) at (1.4,-0.1) {};
		\node[cstate] (s1011) at (0,-1.1) {$\St{1,10,1,1}$};
		\node[cstate] (s0921) at (3,1)    {$\St{0,9,2,1}$};
		\node[cstate] (s0912) at (3.4,0.1)  {$\St{0,9,1,2}$};
			\node[dot] (s0912n) at (4.6,0.1) {};
		\node[cstate] (s0813) at (6,0.9)  {$\St{0,8,1,3}$};
		\node[cstate] (s0712) at (6.2,-0.1) {$\St{0,7,1,2}$};
		\node[cstate] (s1712) at (6.2,-1.2)   {$\St{1,7,1,2}$};
		\node[cstate] (s0811) at (2.5,-1.1) {$\St{0,8,1,1}$};
		
		\path[-,shorten >=0pt]
			(s0011.east) edge[ ] node[pAction,above left,pos=.8] {$\kwm{E}$} (s0011e)
			(s0011.east) edge[ ] node[pAction,above,pos=.8] {$\kwm{N}$} (s0011n)
			(s0912.east) edge[ ] node[pAction,above,pos=.8] {$\kwm{N}$} (s0912n)
			;
		\path[->]
			(s0011)  edge[ ] node[pAction,right] {$\kwm{sleep}$} (s1011)
			(s0712)  edge[ ] node[pAction,right] {$\kwm{sleep}$} (s1712)
			(s0011n) edge[bend left=10] node[pDistr,above,pos=0.7] {$0.9$} (s0912)
			(s0011n) edge[bend left=10] node[pDistr,below left] {$0.1$} (s0811)
			(s0011e) edge[bend left=10] node[pDistr,above left] {$0.9$} (s0921)
			(s0011e) edge[bend left=10] node[pDistr,above right,pos=0.8,outer sep=-2pt] {$0.1$} (s0811)
			(s0912n) edge[bend left=10] node[pDistr,above left,pos=.8] {$0.9$} (s0813.west)
			(s0912n) edge[bend left=10] node[pDistr,below left,pos=.8] {$0.1$} (s0712.west)
			;
		\path[->,out=10,in=-10,looseness=4]
			(s1011) edge (s1011)
			(s1712) edge (s1712)
			;

		\dashedactions{s0921.east}
		\dashedactions{s0811.east}
		\dashedactions{s0712.east}
		\dashedactions{s0813.east}

	\end{scope}
	\end{tikzpicture}}
	\label{fig:example:b}}
	\caption{A motivating example of a robot (left) and part of its model (right).}
	\label{fig:example}
\end{figure}
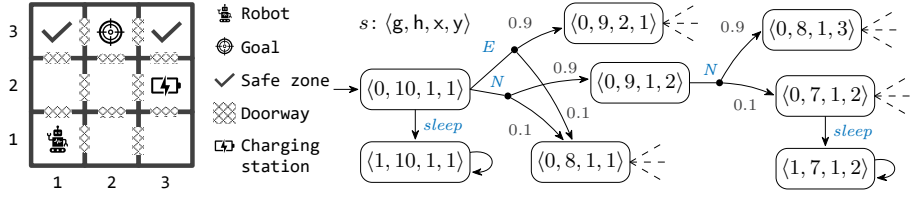 

Now let us examine how such requirements can be formalized.
Let
$ \varphi_{A} =  \Eventually ( \LGoal \land (h\!>\!3) \land \LOn ) $,
$ \varphi_{B} =  \Eventually ( \LChrg \land (h\!>\!3) \land \LOn ) $, and
$ \varphi_{C} =  \Eventually ( \LErr ) $.
One can use PCTL to capture each objective separately as the reachability queries
$ \Phi_{A} = \Pquery_{\max} [ \varphi_{A} ] $,
$ \Phi_{B} = \Pquery_{\max} [ \varphi_{B} ] $, and
$ \Phi_{C} = \Pquery_{\max} [ \varphi_{C} ] $.
A multi-objective query $\Phi_1 = \Phi_{A} \lor \Phi_{B} \lor \Phi_{C} $ does not capture
 the underlying priority structure in the original requirements.
In fact, an optimal strategy for $\Phi_1$
 always chooses the actions that reflect the objective with the highest probability
 of success, resulting in a strategy where the robot simply signals an error from the
 very initial state.
\HL{Similarly, the use of}
$\Phi_2 = \Pquery_{\max} [ \varphi_{A} \Weakuntil \varphi_{B} ]$
 does not provide means to specify
 the context upon which switching from $\varphi_{A}$ to $\varphi_{B}$ occurs.
Attempts featuring multi-objective queries with nested operators, such as
$ \Phi_3 =  \Pquery_{\max} [ \varphi_{A} \land \Pquery_{\max \geq 0.8} [\varphi_{A} ] ] 
 \lor \Pquery_{\max} [ \varphi_{B} \land \Pquery_{\max <0.8} [\varphi_{A} ] ] $,
 have several drawbacks.
First, correctly formalizing the requirement is typically cumbersome and 
 hard to troubleshoot.
Second, to the best of our knowledge, nested queries in the form of 
 $\Pquery_{\opt \in J}$ are not supported by model checkers.
Third, the semantics of the formalized requirement is potentially different from 
 the original one.
For instance,
 $\Phi_3 $ allows the system to pursue $\varphi_{A}$ even after switching to $\varphi_{B}$ 
 if the probability of achieving $\varphi_{A}$ rises again above $0.8$ ---
 a behavior that was not called for in the original requirement.

\HL{Consequently,} in this paper we focus on two problems:
 the formalization of PCTL objectives with an underlying context-based
 priority structure, and 
 the synthesis of strategies for such objectives.
The first problem is addressed by introducing CAPTL in~\secref{sec:logic},
while the second is addressed in~\secref{sec:synthesis}.
We will use this motivating example as a running one throughout the rest of this~paper.

\section{Context-Aware Temporal Logic}
\label{sec:logic}

\subsubsection{CAPTL Syntax.}

CAPTL features two pertinent notions, namely, objectives and contexts.
Let $\MDP$ be our system model, and let $\TaskSet$ be the set of all possible
 PCTL path formulas defined for $\MDP$.
In CAPTL, we define an \emph{objective} $q$ as a conjunctive optimization query 
$
	q = \bigwedge_{i=1}^{m} \Pquery_\Opt \left[ \varphi_i \right] , \;
	\varphi_i \in \TaskSet , \; m > 0 .
$
When $ m>1$, $q$ resembles a multi-objective optimization query in the conjunctive form.
Otherwise, in the simplest form where $m=1$, $q$ is a single-objective query.

A \emph{context} $ w_{ \langle q,q^\prime \rangle } $ marks a state where 
 switching from objective $q$ to objective $q^\prime$ is required.
Formally, we define a context $w$ over $\TaskSet$ as a set of satisfaction queries in
 the disjunctive normal form
$
	w = 
	\bigvee_{j=1}^{n} \bigwedge_{i=1}^{m}
	\Pquery_{\Opt \in J_{ij} } \left[ \varphi_{i,j} \right]  , \;
	\varphi_{ij} \in \Xi , \;
	J \subseteq \left[ 0,1 \right] .
$
Intuitively, in a state where $ w_{\langle q,q^\prime \rangle} $ is satisfied, the 
 system switches from $q$ to $q^\prime$.
Notice that the context definition 
 utilizes the operator 
 $\Pquery_{\Opt \in J_{ij} }$ with an interval,
 \ie a context is evaluated at a given state as a boolean value in $\Booleans$.
In contrast, the objective definition 
 utilizes the operator
 $\Pquery_{\Opt}$ without intervals, \ie a quantitative optimization query that 
 can return a numerical value in $[0,1]$.

A \emph{CAPTL requirement} defines a set of objectives to be satisfied,
 in addition to a set of contexts, representing the probabilistic conditions
 upon which objectives are prioritized.
Formally, we define the syntax of a CAPTL requirement as follows.

\begin{definition}[CAPTL Requirement]\label{def:captl_syntax} 
Given a set of PCTL path formulas $\TaskSet$, 
a \emph{CAPTL requirement} is a tuple 
 $\CAPTL = \CAPTLdef $ where
\begin{itemize}
\item $ Q \subset \left\lbrace \bigwedge_{i=1}^{m} \Pquery_\Opt \left[ \varphi_i \right]
	\mid \varphi_i \in \TaskSet \right\rbrace $
	is a finite nonempty set of objectives over $\TaskSet$,
\item $ W \!\subset\! \left\lbrace \bigvee_{j=1}^{n} \bigwedge_{i=1}^{m}
	\Pquery_{\Opt \in J_{ij} } \left[ \varphi_{i,j} \right]  \mid
	\varphi_{ij} \in \Xi , J_{ij} \subseteq \left[ 0,1 \right] \right\rbrace $
	is a set of contexts,
\item $\hookrightarrow \subseteq Q \times W \times Q $ 
	is a conditional transition relation, and
\item $q_0 \in Q $ is an initial objective.
\end{itemize}
\end{definition}

In a CAPTL requirement $\CAPTL$, each state $q \in Q$ represents an objective,
 \ie an optimization query to be satisfied.
The conditional transition relation $\hookrightarrow $ defines how objectives
 are allowed to change. 
For instance, if {\larger[-1.2]$ q \xhookrightarrow{w} q^\prime $}, 
 a shorthand for $\left( q, w, q^\prime \right) \in \hookrightarrow $,
 then the objectives are switched from $q$ to $q^\prime $ if $w$ is satisfied.
Notice that contexts are used as labels for the conditional transition relation. 
In the rest of this paper, we will overload the notation and use
 $W : Q \rightarrow \Pset(W) $ 
 to denote the set of contexts emerging from a given objective.
We will also use 
 $ Q(q,w) = q' $ 
 to denote that objective 
 $ q $ has a context $ w$ that leads to $ q^\prime $.

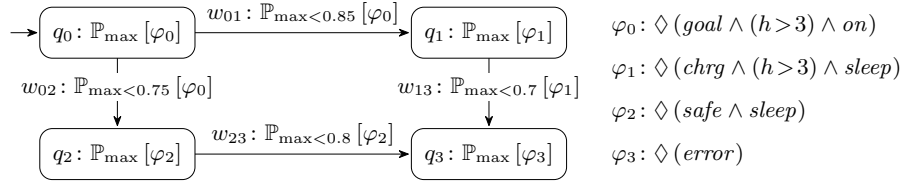
\begin{figure}[t]
	\centering
	\scalebox{0.95}{\begin{tikzpicture}[automaton]
	\def\DistX{5.2}
	\def\DistY{1.7}
	\def\DeltaEq{0.6}
	\begin{scope}[shift={(0,0)}]
		\node[objective,initial] (q0) at (0,0) 
			{$q_0 \colon \Pquery_{\max} \left[ \varphi_0 \right]$};
		\node[objective] (q1) at	(\DistX,0) 
			{$q_1 \colon \Pquery_{\max} \left[ \varphi_1 \right]$};
		\node[objective] (q2) at	(0,-\DistY) 
			{$q_2 \colon \Pquery_{\max} \left[ \varphi_2 \right]$};
		\node[objective] (q3) at	(\DistX,-\DistY) 
			{$q_3 \colon \Pquery_{\max} \left[ \varphi_3 \right]$};
		\path[->]
			(q0) edge[out=0,in=180] node[above,pos=.5]
				{$w_{01} \colon \Pquery_{\max <0.85} \left[ \varphi_0 \right] $} (q1)
			(q2) edge[out=0,in=180] node[above,pos=.5]
				{$w_{23} \colon \Pquery_{\max <0.8} \left[ \varphi_2 \right] $} (q3)
			(q0) edge[out=-90,in=90] node[pos=.4,fill=white,inner sep=1pt,outer sep=0pt]
				{$w_{02} \colon \Pquery_{\max <0.75} \left[ \varphi_0 \right] $} (q2)
			(q1) edge[out=-90,in=90] node[pos=.4,fill=white,inner sep=1pt,outer sep=0pt]
				{$w_{13} \colon \Pquery_{\max <0.7} \left[ \varphi_1 \right] $} (q3);
	\end{scope}
	\begin{scope}[shift={(\DistX+1.6,0.1)}]
		\node[anchor=west] (phi0) at (0,0)
			{$\varphi_0 \colon \Eventually 
					\left( \LGoal \land (h\!>\!3) \land \LOn \right) $};
		\node[below=\DeltaEq of phi0.west, anchor=west] (phi1) 
			{$\varphi_1 \colon \Eventually 
					\left( \LChrg \land (h\!>\!3) \land \LSleep \right) $};
		\node[below=\DeltaEq of phi1.west, anchor=west] (phi2) 
			{$\varphi_2 \colon \Eventually \left( \LSafe \land \LSleep \right) $};
		\node[below=\DeltaEq of phi2.west, anchor=west] (phi3) 
			{$\varphi_3 \colon \Eventually \left( \LErr \right) $};
	\end{scope}
	\end{tikzpicture}}\RemoveSpaceAfterTikz
	\caption{The CAPTL requirement for the running example.}
	\label{fig:captl_example}
\end{figure}

\begin{example}\label{ex:syntax} 
For the running example, 
 \figref{fig:captl_example} shows an example of a CAPTL requirement $\CAPTL$
 where
 $ Q = \{ q_0, q_1, q_2, q_3 \} $,
 $ W = \{ w_{01}, w_{02}, w_{13}, w_{23} \}$,
 and 
 $ \hookrightarrow = \{  \mathsmaller{ 
 	\langle q_0, w_{01}, q_1 \rangle,
 	\langle q_0, w_{02}, q_2 \rangle,
 	\langle q_1, w_{13}, q_3 \rangle,
 	\langle q_0, w_{23}, q_3 \rangle} \} $. 
The requirement starts by prioritizing
 $q_0 = \Pquery_{\max} \left[ \varphi_0 \right] $.
\HL{If $\Pquery_{\max} \left[ \varphi_0 \right] \in [0.75, 0.85)$,}
 the context $w_{01}$ becomes true, 
 and by executing
 {\larger[-1.2]$q_0 \xhookrightarrow{w_{01}} q_1$},
 $q_1 = \Pquery_{\max} \left[ \varphi_1 \right] $ is prioritized.
\HL{Similarly, if $\Pquery_{\max} \left[ \varphi_0 \right] \in [0, 0.75)$,}
 $w_{02}$ becomes true, executing 
 {\larger[-1.2]$q_0 \xhookrightarrow{w_{02}} q_2$}
 where 
 $ q_2 = \Pquery_{\max} \left[ \varphi_2 \right] $ is prioritized.
Notice that objectives can have a single context, \eg 
 $W(q_1) = \left\{ w_{13} \right\} $;
 multiple contexts, \eg
 $W(q_0) = \left\{ w_{01}, w_{02} \right\} $;
 or none, \eg $ W \left(q_3 \right) = \emptyset $.
\end{example}

\subsubsection{CAPTL Semantics for MDPs.}

We progressively define CAPTL semantics for MDPs by first defining 
 the satisfaction semantics for objectives and contexts. 
Let $q = \Pquery_{\max} \left[ \varphi \right]$ be the objective at state $s$,
 and let $\Strset $ be the set of all strategies for $\MDP$.
We say that $s, \Str^* \models q$ if $\Str^* \in \Strset$ such that
\begin{align}\label{eq:q_satisfaction}
	\Prob_\MDP^{\Str^*, s} = \sup_{\Str \in \Strset} \Prob_{\MDP,s}^\Str \left( 
		\left\lbrace \pi \in \Paths_{\MDP,s}^{\Str} \mid \pi \models \varphi \right\rbrace
	\right).
\end{align}
In that case, we call $\Str^*$ a \emph{local strategy}, \ie an optimal strategy 
 w.r.t. $\langle q,s \rangle$.
\begin{definition}[Local Strategy]\label{def:local_strategy}
Let $q_i = \Pquery_\Opt \left[ \varphi_i \right]$ be an objective.
A \emph{local (optimal) strategy} for $\left\langle q_i, s_i \right\rangle$
 is a strategy $ \Str_{\left\langle q_i, s_i \right\rangle} \in \Strset$
 such that
\begin{align*}
	\Prob_{\MDP, s_i }^{\Str_{\left\langle q_i, s_i \right\rangle} } 
		= \opt_{\Str \in \Strset} \Prob_{\MDP,s_i}^\sigma \left(
			\left\lbrace \pi \in \Paths_{\MDP,s_i}^{\Str} \mid 
			\pi \models \varphi_i \right\rbrace
		\right) 
\end{align*}
\end{definition}

Next, let $(q, w, q^\prime) \in \hookrightarrow $,
 where $w = \Pquery_{\leq c } \left[ \varphi \right] $.
Let $s_k \in \Reach(\MDP, s, \Str^* ) $, where $\Str^*$ is the local strategy for
 $ \langle q,s \rangle$. 
We say that $s_k \models w $ if
\begin{align}\label{eq:w_satisfaction}
	\sup_{\Str \in \Strset} \Prob^{\Str}_{\MDP,s_k} \left( 
		\left\lbrace \pi \in \Paths_{\MDP,s_k}^{\Str} \mid 
		\pi \models \varphi \right\rbrace
	\right) \leq c. 
\end{align}
Note that contrary to \eqref{eq:q_satisfaction}, 
 the set of paths $ \{ \pi \} $ in
 \eqref{eq:w_satisfaction} is \emph{not} limited to those induced by the local
 strategy $\Str^* $.
Moreover, if $\exists \pi = s \ldots s_i \ldots s_k \in \Fpath_{\MDP,s}^{\Str} $
 s.t. $s_i \models w $,
 and $s_i \not\models w$ for all $ i < k $, then $s_k$ is called
 a \emph{switching state},
 \ie the first state on a path $\pi$ to satisfy $w$,
 triggering a switch from $q$ to $q^\prime$.

\begin{definition}[Switching Set]\label{def:switch} 
Let $q = \Pquery_\Opt \left[ \varphi \right]$ and $\Str^* \in \Strset$ such that
 \HL{$s_0,\Str^* \models q$.}
The \HL{corresponding} \emph{switching set} 
 $S_q \subseteq \Reach(\MDP,s_0,\Str^*) $ is defined as
\HL{\begin{align*}
	S_{q} = \bigg\lbrace s_k \mid
	\exists \pi = s_0 \ldots s_i \ldots s_k \in \Fpath_{\MDP,s_0}^{\Str^*}  
	\;\suchthat\; 
	s_i \not\models \bigvee_{\mathclap{w \in W(q)}} w ,\, \forall i \!<\! k ;\,
	s_k \models \bigvee_{\mathclap{w \in W(q)}} w \,
	 \bigg\rbrace . 
\end{align*}}
We use $S_q^{q^\prime}$ to denote the set of switching states from $q$ to $q^\prime$.
\end{definition}

An objective is \emph{active} in a state $s$ if it is being pursued at that state.

\begin{definition}[Active Objective]\label{def:active} 
Let
 $\CAPTL = \CAPTLdef $
 and $\MDP = \MDPdef $.
An \emph{activation function} $g \!:\! S \rightarrow \Pset(Q) $ is defined inductively as:
	\HL{(i)} $ g(s_0) \ni q_0 $;
	\HL{and (ii)} $ g(s) \ni q^\prime $ if $g(s) \ni q $ and $s \in S_{q}^{q^\prime} $.
We say objective $q \in Q$ is \emph{active} at state $s \in S$ if $g(s) \ni q$.
\end{definition}

As \HL{captured} in~\defref{def:local_strategy}, local strategies 
are tied to their
 respective objectives.
Consequently, a local strategy is switched whenever an objective is switched as well,
 and the new local strategy substitutes its predecessor.
We call the set of local strategies a \emph{strategy profile}, and the resulting 
 behavior a \emph{protocol}.

\begin{definition}[Protocol]\label{def:protocol} 
Let
 $\CAPTL = \CAPTLdef $
 and $\MDP = \MDPdef $.
Given a strategy profile 
 $ \Str = \left\{ \Str_{\langle q,s \rangle} \ldots \right\} $,
 the induced \emph{(optimal) protocol} is a (partial) function 
 $\Protocol : Q \times S \nrightarrow \Act \cup \Pset(W) $
 such that 
\begin{itemize}
\item $ \Protocol(q,s) = \Str_{ \langle q,s \rangle }(s) \in \Act $ iff 
	$q \in g(s)$ and $ s \not\in S_q $; and
\item $ \Protocol(q,s) \ni w_{\langle q,q^\prime \rangle } \in W $ iff 
	$q \in g(s)$ and $ s \in S_q^{q^\prime} $.
\end{itemize}
\end{definition}

Given $\langle q,s \rangle $, a protocol assigns either an optimal
 action based on the local strategy associated with $q$,
 or a context to switch the active objective itself.
We will use $\Protocolset$ to denote the set of all possible protocols.

\begin{definition}[System-Protocol Composition]
Let $\MDP = \MDPdef $ and 
 \HL{$\Protocol : Q \times S \nrightarrow \Act \cup \Pset(W) $}
 be
 a compatible protocol.
Their composition is defined as 
\HL{%
 $\MDP^\Protocol = \left(\hat{Q}, \Act \cup W, \hat{\Pmatrix}, \hat{s}_0, \hat{L} \right)$
 where 
 $\hat{Q} \subseteq Q \times S $, $\hat{s}_0 = \St{q_0,s_0}$, and}
\HL{\begin{align*}
\hat{\Pmatrix} \left( \St{q,s}, a, \St{q',s'} \right) = \left\{
\begin{array}{lllll}
	\Pmatrix \left( s,a, s' \right) & \mbox{ if } \, \Protocol(q,s) = a ,\, q'=q , \\
	1 & \mbox{ if } \, \Protocol(q,s) = w ,\, s'=s ,\, q'=Q(q,w) , \\
	0 & \mbox{ otherwise} .
\end{array} \right.
\end{align*}}
\end{definition}

We now define the CAPTL satisfaction semantics as follows.

\begin{definition}[CAPTL Satisfaction Semantics]\label{def:captl_semantics} 
Let
 $\CAPTL = \CAPTLdef $,
 $\MDP = \MDPdef $,
 and
 $\Protocol : Q \times S  \nrightarrow \Act \cup \Pset(W) $.
The \emph{CAPTL satisfaction semantics} is defined inductively
 as follows:
\begin{align*}\renewcommand*{\arraystretch}{1.2}
\begin{array}{lllll}
\MDP, \Protocol  & \models & q & \iff &
	\Probability^{\max}_{\MDPxPRO} \left(\left\{ \pi \in \Paths_{\MDPxPRO} \mid
	\Last(\pi)\!=\!\St{q,s'} , s' \models q \right\}\right) \geqslant 1 , \\
\MDP, \Protocol  & \HL{\models_c} & \CAPTL & \iff &
	\Probability^{\max}_{\MDPxPRO} \left(\left\{ \pi \in \Paths_{\MDPxPRO} \mid
	\Last(\pi)\!=\!\St{q,s'} , s' \models q, q\in Q \right\}\right) \HL{= c} , \\
\MDP, \Protocol  & \models & \CAPTL & \iff &
	\MDP, \Protocol  \models_{\geqslant 1} \CAPTL .
\end{array}
\end{align*}
\end{definition}

CAPTL semantics dictate that $\MDP$ and $\Pi $ satisfy $\CAPTL$ if every path
 $\pi \in \Paths_\MDPxPRO $ ends with a state $s \in S$ where $ q \ni g(s)$ and
 $ s \models q $,
 \ie the system reaches some state $s$ where some objective $q$ is both active 
 and satisfied.

\subsubsection{CAPTL Fragments.}

A CAPTL requirement is \emph{nondeterministic} if for some $q\in Q$,
 $\exists w_i, w_j \in W(q) $ such
 that $S_q^{q_i} \cap S_q^{q_j} \neq \emptyset $.
That is, at least one objective has two or more contexts that can be active at the same
 state.
\HL{If that is not the case,}
then the CAPTL requirement is \emph{deterministic}.
We now identify a fragment of deterministic CAPTL requirements 
 where \HL{the following} two conditions are met.
First, every $ q \in Q$ is a quantitative PCTL persistence objective.
Second, every $ w \in W(q) $ is a qualitative PCTL persistence objective
 over the same persistence set as in $q$.
This is formally captured in the following definition.
\begin{definition}[Persistence CAPTL]\label{def:normal_form} 
A CAPTL requirement $\CAPTL = \CAPTLdef $ is
 \emph{persistent} 
 if every $q \in Q$ is of the form
 $q = \Pquery_{\max } [\Eventually\Always B] $; and
 if $ W(q) \neq \emptyset $ then $\forall w_{\langle q, q_j \rangle} \in W(q)$,
 	$ w_{\langle q, q_j \rangle} = 
 	\Pquery_{\max \in J_{j} } [\Eventually\Always B] $
 	such that \HL{$(J_{j})$} are disjoint intervals where 
 	\HL{$\cup_j J_{j} = \left[0,c\right) $, $0 < c \leq 1$.}
\end{definition}

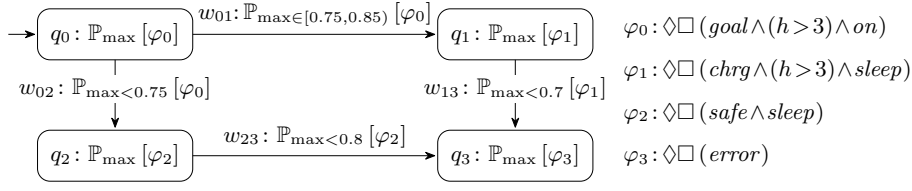
\begin{figure}[t]
	\centering
	\scalebox{0.95}{\begin{tikzpicture}[automaton]
	\def\DistX{5.6}
	\def\DistY{1.7}
	\def\DeltaEq{0.6}
	\begin{scope}[shift={(0,0)}]
		\node[objective,initial] (q0) at (0,0) 
			{$q_0 \colon \Pquery_{\max} \left[ \varphi_0 \right]$};
		\node[objective] (q1) at	(\DistX,0) 
			{$q_1 \colon \Pquery_{\max} \left[ \varphi_1 \right]$};
		\node[objective] (q2) at	(0,-\DistY) 
			{$q_2 \colon \Pquery_{\max} \left[ \varphi_2 \right]$};
		\node[objective] (q3) at	(\DistX,-\DistY) 
			{$q_3 \colon \Pquery_{\max} \left[ \varphi_3 \right]$};
		\path[->]
			(q0) edge[out=0,in=180] node[above,pos=.5]
				{$w_{01} \!\colon\! \Pquery_{\max \in \left[0.75,0.85 \right) } 
					\left[ \varphi_0 \right] $} (q1)
			(q2) edge[out=0,in=180] node[above,pos=.5]
				{$w_{23} \colon \Pquery_{\max <0.8} \left[ \varphi_2 \right] $} (q3)
			(q0) edge[out=-90,in=90] node[pos=.4,fill=white,inner sep=1pt,outer sep=0pt]
				{$w_{02} \colon \Pquery_{\max <0.75} \left[ \varphi_0 \right] $} (q2)
			(q1) edge[out=-90,in=90] node[pos=.4,fill=white,inner sep=1pt,outer sep=0pt]
				{$w_{13} \colon \Pquery_{\max <0.7} \left[ \varphi_1 \right] $} (q3);
	\end{scope}
	\begin{scope}[shift={(\DistX+1.4,0.1)}]
		\node[anchor=west] (phi0) at (0,0)
			{$\varphi_0 \colon\! \Eventually\Always 
					\left( \LGoal \!\land\! (h\!>\!3) \!\land\! \LOn \right) $};
		\node[below=\DeltaEq of phi0.west, anchor=west] (phi1) 
			{$\varphi_1 \colon\! \Eventually\Always 
					\left( \LChrg \!\land\! (h\!>\!3) \!\land\! \LSleep \right) $};
		\node[below=\DeltaEq of phi1.west, anchor=west] (phi2) 
			{$\varphi_2 \colon\! \Eventually\Always \left( \LSafe \!\land\! \LSleep \right) $};
		\node[below=\DeltaEq of phi2.west, anchor=west] (phi3) 
			{$\varphi_3 \colon\! \Eventually\Always \left( \LErr \right) $};
	\end{scope}
	\end{tikzpicture}}\RemoveSpaceAfterTikz
	\caption{The persistence CAPTL requirement for the running example.} 
	\label{fig:ex:normal_form}
\end{figure}

A persistence CAPTL (P-CAPTL) requirement allows for defining persistence objectives,
 where each objective maximizes the probability of 
 (\ie prioritizes) reaching a corresponding persistence set.
Contexts in this case can be understood as lower bounds of their respective objectives.
That is, an objective is pursued as long as, at any transient state, the probability
 of achieving such objective does not drop below a certain threshold.
The requirement also ensures that at most one context is satisfied at any state,
 eliminating any nondeterministism in $\CAPTL$.
\begin{example}\label{ex:normal_form} 
Continuing~\exref{ex:syntax}, 
 \figref{fig:ex:normal_form} shows the persistence CAPTL requirement for the robot.
Notice that all objectives are in the form 
 $\Pquery_{\max} [\Eventually\Always B ] $.
Also, the intervals \HL{$[0.75,0.85)$ and $[0,0.75)$}
 of $w_{01} $ and $w_{02}$, respectively,
 are disjoint, hence at most one context 
 in $W(q_0) = \left\{ w_{01}, w_{02} \right\} $ can be satisfied at any state.
\end{example}

\section{CAPTL-Based Synthesis}
\label{sec:synthesis}

In this section we first define the synthesis problem for CAPTL requirements.
Next, we examine a general procedure for deterministic CAPTL where the synthesis 
 problem is reduced to \HL{solving a set} of PCTL-based strategy synthesis problems.
Finally, we utilize the underlying structure of persistence properties
 to propose a synthesis procedure optimized for P-CAPTL requirements.

\subsubsection{CAPTL Synthesis Problem.}

In the rest of this section, let
 $\MDP = \MDPdef$ and 
 $\CAPTL = \CAPTLdef$.
We assume that a probabilistic model checker is given 
 (\eg \PRISMG{}~\cite{kwiatkowska2018prism} or \UPPAALS{}~\cite{david2015uppaal})
 that can accept an MDP-based model $\MDP$ and a PCTL formula $\Phi$ as inputs,
 and provides the following functions:
\begin{itemize}
\item $ \Reachset \!::\! \left(\MDP, s \right) \mapsto R \subseteq S $
	returns $ R = \Reach(\MDP, s) $. 
\item $ \Verify \!::\! \left(\MDP, s, \Phi \right) \mapsto b \in \Booleans $
	returns $\top$ iff $\MDP,s \models \Phi$, and $\bot$ otherwise.
\item $ \Synth \!::\! \left(\MDP, s, \Phi \right) \mapsto \left( \Str, c \right)  $
	finds $\Str \in \Strset $ s.t. 
	$\Prob \left( \MDP_s^\Str \models \Phi \right) = c \in [0,1] $.
\end{itemize}
We also assume that the model checker functions terminate in finite time and return 
 correct answers.
We now define the CAPTL synthesis problem as follows.

\begin{definition}[CAPTL Synthesis Problem] 
Given 
 $\MDP = \MDPdef $ and
 $\CAPTL = \CAPTLdef $,
 the \emph{CAPTL synthesis problem} seeks to find a protocol
 $\Protocol : Q \times S \nrightarrow \Act \cup W $ such that
 $\MDP, \Protocol \models \CAPTL $.
\end{definition}

\subsubsection{PCTL-Based Approach.}

The synthesis problem can be reduced to solving a set of PCTL-based synthesis queries as
 demonstrated in~\algref{alg:pctl}.
Starting with $\St{q_0,s_0}$, the algorithm verifies whether any context 
 $w \in W(q_0)$ is satisfied, and if true, adds $w$ to the protocol and switches 
 to the next objective.
If no context is satisfied, the algorithm synthesizes a local strategy and adds
 the corresponding optimal action to the protocol.

\begin{algorithm}[!tb] 
\caption{PCTL-Based Synthesis} \label{alg:pctl}
\KwIn{$\MDP = \MDPdef $, $\CAPTL = \CAPTLdef $}
\KwResult{$\Protocol, c$ such that $\MDP, \Protocol \models_c \CAPTL$}
	\lForEach(){$q\in Q$}{%
		$\hat{S}_{q} \assign \emptyset$,\,
		$\bar{S}_{q} \assign \emptyset$
	}	
	$ \Protocol \assign \emptyset $,\, 	
	$ \hat{S}_{q_0} \assign \left\{ s_0 \right\} $,\, 
	$ q \assign q_0 $,\,
	$ \vec{C} \assign \vec{0}_{Q \times S}\in [0,1]^{Q \times S} $, \, 
	$ \VRepeat \assign \top $ \;
	\While(){$\hat{S}_{q} \neq \emptyset $}{\label{alg:pctl:firstwhile}
		Let $ s \in \hat{S}_{q}$,\,
		$ \hat{S}_{q} \assign \hat{S}_{q} \setminus \{ s \} $,
		$\bar{S}_{q} \assign \bar{S}_{q} \cup \{ s \} $\;\label{alg:pctl:sremove}
		\While($ \mathit{repeat} \assign \bot $){$ \mathit{repeat} $}{%
			\label{alg:pctl:secondwhile}
			\ForEach(){ $ w \in W(q) $ }{\label{alg:pctl:foreach}
				\If(){ $ \Verify \left( \MDP, s, w \right) = \top  $}{
					$ \Protocol \assign \Protocol \cup \{ (s, q, w)  \} $,\,
					$ q \assign Q(q, w) $,\,
					$\mathit{repeat} \assign \top $,\,
				  \label{alg:pctl:qupdate}
					\Break 
				\RemoveAlgExtraSpace}
			\RemoveAlgExtraSpace}
		\RemoveAlgExtraSpace}
		$ \left(\Str, \vec{C}(q,s) \right) \assign \Synth \left( \MDP; s, q \right) $,\,
		$ \Protocol \assign \Protocol \cup \{ \left( s, q, \Str(s) \right)  \} $ \;
		$ \hat{S}_{q} \assign \hat{S}_{q} \cup 
			\left(\Post \left(\MDP, s,\Str(s) \right) \setminus \bar{S}_{q} \right) $
			\label{alg:pctl:post}
	\RemoveAlgExtraSpace}
	$ c \assign \Verify \left( \MDPxPRO,\, \St{q_0,s_0} ,\, 
		\Pquery \left[ \Eventually \bigvee_{q\in Q} (\St{q,s} \land \vec{C}(q,s) \!=\! 1)
		\right] \right) $
\end{algorithm}

\begin{proposition}\label{pro:pctl} 
\algref{alg:pctl} terminates; and returns $\Protocol, c$ iff 
 $\MDP,\Protocol \models_c \CAPTL $.
\end{proposition}

\subsubsection{Synthesis for P-CAPTL}

We now propose a synthesis algorithm optimized for persistence CAPTL.
To this end, we show that for a given persistence objective, synthesizing a local
 strategy in the initial state suffices.
In a manner similar to switching states (see~\defref{def:switch}), 
 we devise a partition of reachable states for every objective.
We will use those concepts to define a system-CAPTL composition and show that it
 is bisimilar to $\MDPxPRO$.

Let $ R = \Reach(\MDP,s_0)$.
We first note that given $\MDP$ and $q = \Pquery_\Opt [\Eventually\Always B ]$, 
 existing model checking and synthesis algorithms typically compute
 a least fixed point (LFP) vector $ \vec{x}_q \in \left[0,1 \right]^{ |R| } $, where
 $\vec{x}_q[s]$ is the optimal probability of satisfying $\Eventually\Always B $ at state
 $s \in R $ \HL{(\eg see \cite{baier2008principles,kattenbelt2010game}).}
That is, when 
 $\Synth\left( \MDP, s_0, q \right) $ is called,  $\vec{x}_q$ is computed,
 but only $c=\vec{x}_q[s_0]$ is returned
 (\ie the value at the initial state).
We exploit this fact by implementing a function
 $ \ReachPset :: \left(\MDP, s, q \right) \mapsto \vec{x}_q $
 that returns the LFP vector $\vec{x}_q$ associated with $q$.

\begin{lemma}[Local Strategy Dominance]\label{lem:lsd} 
Let $ \MDP = \MDPdef $ and $q=\Pquery_{\max} [\Eventually\Always B]$.
For all $ s \in \Reach(\MDP,s_0) $, 
 $\Str_\St{q,s} = \restr{\Str_\St{q,s_0} }{\Reach(\MDP,s)} $.
\end{lemma}

\lemref{lem:lsd} signifies that a local strategy for $q$ in the 
 initial state (\ie $\Str_\St{q,s_0}$) subsumes all local strategies
 for the same probabilistic reachability objective in every $s\in R$.
Next, for every $q \in Q$, let us define the following partition of $R$:
\begin{itemize}
\item 
	$ R_q^q = \left\{ s \in R \mid 
		\forall w = \Pquery_{\max \in J}[\Eventually\Always B] \in W(q),\,
		\vec{x}_{q}[s] \not\in J  \right\} $,
	\ie the states in $R$ where, if $q$ is active, keep pursuing $q$.
\item
	$ R_q^{q'} = \left\{ s \in R \mid 
		\exists w = \Pquery_{\max \in J}[\Eventually\Always B] \in W(q), 
		\vec{x}_{q}[s] \in J,\, Q(q,w)=q' \right\} $,
	i.e., the states in $R$ where, if $q$ is active, switch to $q'$.
\end{itemize}

\begin{lemma}[Partitioning]\label{lem:ssp} 
Let $ \MDP = \MDPdef $, $\CAPTL = \CAPTLdef$, and $R = \Reach(\MDP,s_0)$.
For every $q \in Q$,
 $ \bigcup_{q' \in Q} R_q^{q'} \allowbreak = R $; and
 $ R_q^{q'} \cap R_q^{q''} = \emptyset $ for every $ q' \neq q'' $.
\end{lemma}

\begin{proofsketch}
From \defref{def:normal_form}, the intervals $(J_w)_{w\in W(q)}$
 are \HL{disjoint;} hence $(R_q^{q'})_{q'\neq q} $ are disjoint as well, and that 
 $ R_q^q = R / \left( \bigcup_{q'\neq q} R_q^{q'} \right) $.  
\end{proofsketch}

\begin{example} 
\HL{Returning to the P-CAPTL requirement specified in the running example} 
(see~\figref{fig:ex:normal_form}), \figref{fig:partitioning} depicts the partitioning of
 the state-space based on $q_0$, $q_1$, $q_2$ and $q_3$.
Notice that for any $q \in Q$, the sets 
 $ ( R_{q}^{q'} )_{ q' \in Q} $ are pairwise
 disjoint, where $\cup_{q' \in Q} R_{q}^{q'} = \Reach(\MDP,s_0)$.
For example, $R_{q_0}^{q_0}$, $R_{q_0}^{q_1}$ and $R_{q_0}^{q_2}$ do not intersect, and
 their union spans $R = \Reach(\MDP,s_0)$.
In this case, $R_{q_0}^{q_3} = \emptyset $ since there is no direct context emerging from
 $q_0$ to $q_3$.

\end{example}

\begin{figure}[!tb] 
	\centering
	\newlength{\LNmksz}
	\setlength{\LNmksz}{3.4pt}
	\pgfplotsset{every tick label/.append style={font=\LARGE}}
	\resizebox{0.7\columnwidth}{!}{
%
%
\begin{tikzpicture}

\begin{axis}[%
width=4.521in,
height=0.3in,
at={(0.758in,1.599in)},
scale only axis,
xmin=0,
xmax=34,
ymin=0,
ymax=3,
axis background/.style={fill=white},
axis x line*=bottom,
axis y line*=left,
axis line style={draw=none},
ticks=none,
legend columns=7,
legend style={
	at={(0.5,0.5)}, anchor=center, 
	legend cell align=left, align=left, draw=white!15!black}
]

\addlegendentry{$S \setminus R \quad $}
\addlegendimage{only marks, mark=square*, mark options={},
				mark size=\LNmksz, draw=white!90!black, fill=white!90!black}

\addlegendentry{$R_{q}^{=1} \quad $}
\addlegendimage{only marks, mark=square, mark options={rotate=0},
				mark size=\LNmksz, draw=RoyalBlue, fill=none}
				
\addlegendentry{$R_{q}^{q} \quad $}
\addlegendimage{only marks, mark=+, mark options={rotate=0}, 
				mark size=\LNmksz, draw=RoyalBlue, fill=none}
				
\addlegendentry{$R_{q_0}^{q_1} \quad $}
\addlegendimage{only marks, mark=triangle, mark options={rotate=0}, 
				mark size=\LNmksz, draw=RedOrange, fill=none}
\addlegendentry{$R_{q_0}^{q_2} \quad $}
\addlegendimage{only marks, mark=triangle, mark options={rotate=180}, 
				mark size=\LNmksz, draw=RedOrange, fill=none}
\addlegendentry{$R_{q_1}^{q_3} \quad $}
\addlegendimage{only marks, mark=triangle, mark options={rotate=270}, 
				mark size=\LNmksz, draw=RedOrange, fill=none}
\addlegendentry{$R_{q_2}^{q_3} $}
\addlegendimage{only marks, mark=triangle, mark options={rotate=90}, 
				mark size=\LNmksz, draw=RedOrange, fill=none}

\end{axis}
\end{tikzpicture}%
}\vspace{-6pt}\\
	\resizebox{0.49\columnwidth}{!}{
%
%
\definecolor{mycolor1}{rgb}{0.49400,0.18400,0.55600}%
\begin{tikzpicture}

\begin{axis}[%
width=4.521in,
height=1.33in,
at={(0.758in,1.599in)},
scale only axis,
xmin=0,
xmax=34,
ymin=0,
ymax=10,
axis background/.style={fill=white},
axis x line*=bottom,
axis y line*=left,
axis line style={draw=none},
xlabel style={font=\color{white!15!black},yshift=20pt}, xlabel={\relax},
ylabel style={font=\color{white!15!black},yshift=-40pt}, ylabel={\relax},
xmajorticks=false,
ytick={5},
yticklabels={$q_3$},
y tick label style={rotate=90},
legend style={
	at={(0.97,0.03)}, anchor=south east, 
	legend cell align=left, align=left, draw=white!15!black}
]

\addplot [draw=none, fill=white!15!black, forget plot] coordinates {(0, 9.3+9*0 ) (0, 0.7+9*0 ) (0.2, 0.7+9*0 ) (0.2, 9.3+9*0 )};

\addplot[only marks, mark=square*, mark options={}, mark size=\LNmksz, draw=white!90!black, fill=white!90!black] table[row sep=crcr]{%
x	y\\
1	1\\
1	2\\
1	3\\
1	4\\
1	5\\
1	6\\
1	7\\
1	8\\
1	9\\
2	1\\
2	2\\
2	3\\
2	4\\
2	5\\
2	6\\
2	7\\
2	8\\
2	9\\
3	1\\
3	2\\
3	3\\
3	4\\
3	5\\
3	6\\
3	7\\
3	8\\
3	9\\
4	1\\
4	2\\
4	3\\
4	4\\
4	5\\
4	6\\
4	7\\
4	8\\
4	9\\
5	1\\
5	2\\
5	3\\
5	4\\
5	5\\
5	6\\
5	7\\
5	8\\
5	9\\
6	1\\
6	2\\
6	3\\
6	4\\
6	5\\
6	6\\
6	7\\
6	8\\
6	9\\
7	1\\
7	2\\
7	3\\
7	4\\
7	5\\
7	6\\
7	7\\
7	8\\
7	9\\
8	1\\
8	2\\
8	3\\
8	4\\
8	5\\
8	6\\
8	7\\
8	8\\
8	9\\
9	1\\
9	2\\
9	3\\
9	4\\
9	5\\
9	6\\
9	7\\
9	8\\
9	9\\
10	1\\
10	2\\
10	3\\
10	4\\
10	5\\
10	6\\
10	7\\
10	8\\
10	9\\
11	1\\
11	2\\
11	3\\
11	4\\
11	5\\
11	6\\
11	7\\
11	8\\
11	9\\
12	1\\
12	2\\
12	3\\
12	4\\
12	5\\
12	6\\
12	7\\
12	8\\
12	9\\
13	1\\
13	2\\
13	3\\
13	4\\
13	5\\
13	6\\
13	7\\
13	8\\
13	9\\
14	1\\
14	2\\
14	3\\
14	4\\
14	5\\
14	6\\
14	7\\
14	8\\
14	9\\
15	1\\
15	2\\
15	3\\
15	4\\
15	5\\
15	6\\
15	7\\
15	8\\
15	9\\
16	1\\
16	2\\
16	3\\
16	4\\
16	5\\
16	6\\
16	7\\
16	8\\
16	9\\
17	1\\
17	2\\
17	3\\
17	4\\
17	5\\
17	6\\
17	7\\
17	8\\
17	9\\
18	1\\
18	2\\
18	3\\
18	4\\
18	5\\
18	6\\
18	7\\
18	8\\
18	9\\
19	1\\
19	2\\
19	3\\
19	4\\
19	5\\
19	6\\
19	7\\
19	8\\
19	9\\
20	1\\
20	2\\
20	3\\
20	4\\
20	5\\
20	6\\
20	7\\
20	8\\
20	9\\
21	1\\
21	2\\
21	3\\
21	4\\
21	5\\
21	6\\
21	7\\
21	8\\
21	9\\
22	1\\
22	2\\
22	3\\
22	4\\
22	5\\
22	6\\
22	7\\
22	8\\
22	9\\
23	1\\
23	2\\
23	3\\
23	4\\
23	5\\
23	6\\
23	7\\
23	8\\
23	9\\
24	1\\
24	2\\
24	3\\
24	4\\
24	5\\
24	6\\
24	7\\
24	8\\
24	9\\
25	1\\
25	2\\
25	3\\
25	4\\
25	5\\
25	6\\
25	7\\
25	8\\
25	9\\
26	1\\
26	2\\
26	3\\
26	4\\
26	5\\
26	6\\
26	7\\
26	8\\
26	9\\
27	1\\
27	2\\
27	3\\
27	4\\
27	5\\
27	6\\
27	7\\
27	8\\
27	9\\
28	1\\
28	2\\
28	3\\
28	4\\
28	5\\
28	6\\
28	7\\
28	8\\
28	9\\
29	1\\
29	2\\
29	3\\
29	4\\
29	5\\
29	6\\
29	7\\
29	8\\
29	9\\
30	1\\
30	2\\
30	3\\
30	4\\
30	5\\
30	6\\
30	7\\
30	8\\
30	9\\
31	1\\
31	2\\
31	3\\
31	4\\
31	5\\
31	6\\
31	7\\
31	8\\
31	9\\
32	1\\
32	2\\
32	3\\
32	4\\
32	5\\
32	6\\
32	7\\
32	8\\
32	9\\
33	1\\
33	2\\
33	3\\
33	4\\
33	5\\
33	6\\
33	7\\
33	8\\
33	9\\
};
\addlegendentry{data1}

\addplot[only marks, mark=square*, mark options={}, mark size=\LNmksz, draw=white, fill=white] table[row sep=crcr]{%
x	y\\
31	1\\
31	2\\
31	3\\
32	1\\
32	2\\
32	3\\
33	1\\
33	2\\
33	3\\
28	1\\
28	2\\
28	3\\
29	1\\
29	2\\
29	3\\
30	1\\
30	2\\
30	3\\
25	1\\
25	2\\
25	3\\
26	1\\
26	2\\
26	3\\
27	1\\
27	2\\
27	3\\
22	1\\
22	2\\
22	3\\
23	1\\
23	2\\
23	3\\
24	1\\
24	2\\
24	3\\
19	1\\
19	2\\
19	3\\
20	1\\
20	2\\
20	3\\
21	1\\
21	2\\
21	3\\
16	1\\
16	2\\
16	3\\
17	1\\
17	2\\
17	3\\
18	1\\
18	2\\
18	3\\
13	1\\
13	2\\
13	3\\
14	1\\
14	2\\
14	3\\
15	1\\
15	2\\
15	3\\
10	1\\
10	2\\
10	3\\
11	1\\
11	2\\
11	3\\
12	1\\
12	2\\
7	1\\
7	2\\
7	3\\
8	1\\
8	2\\
9	1\\
4	1\\
4	2\\
5	1\\
1	1\\
28	4\\
28	5\\
28	6\\
29	4\\
29	5\\
29	6\\
30	4\\
30	5\\
30	6\\
25	4\\
25	5\\
25	6\\
26	4\\
26	5\\
26	6\\
27	4\\
27	5\\
27	6\\
22	4\\
22	5\\
22	6\\
23	4\\
23	5\\
23	6\\
24	4\\
24	5\\
24	6\\
19	4\\
19	5\\
19	6\\
20	4\\
20	5\\
20	6\\
21	4\\
21	5\\
21	6\\
16	4\\
16	5\\
16	6\\
17	4\\
17	5\\
17	6\\
18	4\\
18	5\\
18	6\\
13	4\\
13	5\\
13	6\\
14	4\\
14	5\\
14	6\\
15	4\\
15	5\\
15	6\\
10	4\\
10	5\\
10	6\\
11	4\\
11	5\\
11	6\\
12	4\\
12	5\\
7	4\\
7	5\\
7	6\\
8	4\\
8	5\\
9	4\\
4	4\\
4	5\\
5	4\\
1	4\\
31	7\\
31	8\\
31	9\\
32	7\\
32	9\\
33	7\\
33	8\\
33	9\\
28	7\\
28	8\\
28	9\\
29	7\\
29	8\\
29	9\\
30	7\\
30	8\\
30	9\\
25	7\\
25	8\\
25	9\\
26	7\\
26	8\\
26	9\\
27	7\\
27	8\\
27	9\\
22	7\\
22	8\\
22	9\\
23	7\\
23	8\\
23	9\\
24	7\\
24	8\\
24	9\\
19	7\\
19	8\\
19	9\\
20	7\\
20	8\\
20	9\\
21	7\\
21	8\\
21	9\\
16	7\\
16	8\\
16	9\\
17	7\\
17	8\\
17	9\\
18	7\\
18	8\\
18	9\\
13	7\\
13	8\\
13	9\\
14	7\\
14	8\\
14	9\\
15	7\\
15	8\\
15	9\\
10	7\\
10	8\\
10	9\\
11	7\\
11	8\\
11	9\\
12	7\\
12	8\\
7	7\\
7	8\\
7	9\\
8	7\\
8	8\\
9	7\\
4	7\\
4	8\\
5	7\\
1	7\\
};
\addlegendentry{data2}

\addplot[only marks, mark=+, mark options={}, mark size=\LNmksz, draw=RoyalBlue, fill=white] table[row sep=crcr]{%
x	y\\
31	1\\
31	2\\
31	3\\
32	1\\
32	2\\
32	3\\
33	1\\
33	2\\
33	3\\
28	1\\
28	2\\
28	3\\
29	1\\
29	2\\
29	3\\
30	1\\
30	2\\
30	3\\
25	1\\
25	2\\
25	3\\
26	1\\
26	2\\
26	3\\
27	1\\
27	2\\
27	3\\
22	1\\
22	2\\
22	3\\
23	1\\
23	2\\
23	3\\
24	1\\
24	2\\
24	3\\
19	1\\
19	2\\
19	3\\
20	1\\
20	2\\
20	3\\
21	1\\
21	2\\
21	3\\
16	1\\
16	2\\
16	3\\
17	1\\
17	2\\
17	3\\
18	1\\
18	2\\
18	3\\
13	1\\
13	2\\
13	3\\
14	1\\
14	2\\
14	3\\
15	1\\
15	2\\
15	3\\
10	1\\
10	2\\
10	3\\
11	1\\
11	2\\
11	3\\
12	1\\
12	2\\
7	1\\
7	2\\
7	3\\
8	1\\
8	2\\
9	1\\
4	1\\
4	2\\
5	1\\
1	1\\
28	4\\
28	5\\
28	6\\
29	4\\
29	5\\
29	6\\
30	4\\
30	5\\
30	6\\
25	4\\
25	5\\
25	6\\
26	4\\
26	5\\
26	6\\
27	4\\
27	5\\
27	6\\
22	4\\
22	5\\
22	6\\
23	4\\
23	5\\
23	6\\
24	4\\
24	5\\
24	6\\
19	4\\
19	5\\
19	6\\
20	4\\
20	5\\
20	6\\
21	4\\
21	5\\
21	6\\
16	4\\
16	5\\
16	6\\
17	4\\
17	5\\
17	6\\
18	4\\
18	5\\
18	6\\
13	4\\
13	5\\
13	6\\
14	4\\
14	5\\
14	6\\
15	4\\
15	5\\
15	6\\
10	4\\
10	5\\
10	6\\
11	4\\
11	5\\
11	6\\
12	4\\
12	5\\
7	4\\
7	5\\
7	6\\
8	4\\
8	5\\
9	4\\
4	4\\
4	5\\
5	4\\
1	4\\
31	7\\
31	8\\
31	9\\
32	7\\
32	9\\
33	7\\
33	8\\
33	9\\
28	7\\
28	8\\
28	9\\
29	7\\
29	8\\
29	9\\
30	7\\
30	8\\
30	9\\
25	7\\
25	8\\
25	9\\
26	7\\
26	8\\
26	9\\
27	7\\
27	8\\
27	9\\
22	7\\
22	8\\
22	9\\
23	7\\
23	8\\
23	9\\
24	7\\
24	8\\
24	9\\
19	7\\
19	8\\
19	9\\
20	7\\
20	8\\
20	9\\
21	7\\
21	8\\
21	9\\
16	7\\
16	8\\
16	9\\
17	7\\
17	8\\
17	9\\
18	7\\
18	8\\
18	9\\
13	7\\
13	8\\
13	9\\
14	7\\
14	8\\
14	9\\
15	7\\
15	8\\
15	9\\
10	7\\
10	8\\
10	9\\
11	7\\
11	8\\
11	9\\
12	7\\
12	8\\
7	7\\
7	8\\
7	9\\
8	7\\
8	8\\
9	7\\
4	7\\
4	8\\
5	7\\
1	7\\
};
\addlegendentry{data3}

\addplot[only marks, mark=square, mark options={}, mark size=\LNmksz, draw=RoyalBlue] table[row sep=crcr]{%
x	y\\
28	1\\
28	2\\
28	3\\
29	1\\
29	2\\
29	3\\
30	1\\
30	2\\
30	3\\
25	1\\
25	2\\
25	3\\
26	1\\
26	2\\
26	3\\
27	1\\
27	2\\
27	3\\
22	1\\
22	2\\
22	3\\
23	1\\
23	2\\
23	3\\
24	1\\
24	2\\
24	3\\
19	1\\
19	2\\
19	3\\
20	1\\
20	2\\
20	3\\
21	1\\
21	2\\
21	3\\
16	1\\
16	2\\
16	3\\
17	1\\
17	2\\
17	3\\
18	1\\
18	2\\
18	3\\
13	1\\
13	2\\
13	3\\
14	1\\
14	2\\
14	3\\
15	1\\
15	2\\
15	3\\
10	1\\
10	2\\
10	3\\
11	1\\
11	2\\
11	3\\
12	1\\
12	2\\
7	1\\
7	2\\
7	3\\
8	1\\
8	2\\
9	1\\
4	1\\
4	2\\
5	1\\
1	1\\
28	4\\
28	5\\
28	6\\
29	4\\
29	5\\
29	6\\
30	4\\
30	5\\
30	6\\
25	4\\
25	5\\
25	6\\
26	4\\
26	5\\
26	6\\
27	4\\
27	5\\
27	6\\
22	4\\
22	5\\
22	6\\
23	4\\
23	5\\
23	6\\
24	4\\
24	5\\
24	6\\
19	4\\
19	5\\
19	6\\
20	4\\
20	5\\
20	6\\
21	4\\
21	5\\
21	6\\
16	4\\
16	5\\
16	6\\
17	4\\
17	5\\
17	6\\
18	4\\
18	5\\
18	6\\
13	4\\
13	5\\
13	6\\
14	4\\
14	5\\
14	6\\
15	4\\
15	5\\
15	6\\
10	4\\
10	5\\
10	6\\
11	4\\
11	5\\
11	6\\
12	4\\
12	5\\
7	4\\
7	5\\
7	6\\
8	4\\
8	5\\
9	4\\
4	4\\
4	5\\
5	4\\
1	4\\
31	7\\
31	8\\
31	9\\
32	7\\
32	9\\
33	7\\
33	8\\
33	9\\
28	7\\
28	8\\
28	9\\
29	7\\
29	8\\
29	9\\
30	7\\
30	8\\
30	9\\
25	7\\
25	8\\
25	9\\
26	7\\
26	8\\
26	9\\
27	7\\
27	8\\
27	9\\
22	7\\
22	8\\
22	9\\
23	7\\
23	8\\
23	9\\
24	7\\
24	8\\
24	9\\
19	7\\
19	8\\
19	9\\
20	7\\
20	8\\
20	9\\
21	7\\
21	8\\
21	9\\
16	7\\
16	8\\
16	9\\
17	7\\
17	8\\
17	9\\
18	7\\
18	8\\
18	9\\
13	7\\
13	8\\
13	9\\
14	7\\
14	8\\
14	9\\
15	7\\
15	8\\
15	9\\
10	7\\
10	8\\
10	9\\
11	7\\
11	8\\
11	9\\
12	7\\
12	8\\
7	7\\
7	8\\
7	9\\
8	7\\
8	8\\
9	7\\
4	7\\
4	8\\
5	7\\
1	7\\
};
\addlegendentry{data4}

\legend{}; 
\end{axis}
\begin{axis}[%
width=4.521in,
height=1.33in,
at={(0.758in,1.599in)},
scale only axis,
xmin=0,
xmax=34,
ymin=0,
ymax=10,
axis x line*=bottom,
axis y line*=right,
axis line style={draw=none},
xlabel style={font=\color{white!15!black},yshift=20pt}, xlabel={\relax},
ylabel style={font=\color{white!15!black},yshift=40pt}, ylabel={\relax},
xmajorticks=false,
ytick={2,5,8},
yticklabels={$\var{g}_0$, $\var{g}_1$, $\var{g}_2$},
y tick label style={rotate=90},
legend style={
	at={(0.97,0.03)}, anchor=south east, 
	legend cell align=left, align=left, draw=white!15!black}
]

\addplot [draw=none, fill=white!15!black, forget plot] coordinates {(34-0.2+0, 3.3+3*0 ) (34-0.2+0, 0.7+3*0 ) (34-0.2+0.2, 0.7+3*0 ) (34-0.2+0.2, 3.3+3*0 )};
\addplot [draw=none, fill=white!15!black, forget plot] coordinates {(34-0.2+0, 3.3+3*1 ) (34-0.2+0, 0.7+3*1 ) (34-0.2+0.2, 0.7+3*1 ) (34-0.2+0.2, 3.3+3*1 )};
\addplot [draw=none, fill=white!15!black, forget plot] coordinates {(34-0.2+0, 3.3+3*2 ) (34-0.2+0, 0.7+3*2 ) (34-0.2+0.2, 0.7+3*2 ) (34-0.2+0.2, 3.3+3*2 )};

\end{axis}
\end{tikzpicture}%
}\hfill%
	\resizebox{0.49\columnwidth}{!}{
%
%
\definecolor{mycolor1}{rgb}{0.30100,0.74500,0.93300}%
\begin{tikzpicture}

\begin{axis}[%
width=4.521in,
height=1.33in,
at={(0.758in,1.599in)},
scale only axis,
xmin=0,
xmax=34,
ymin=0,
ymax=10,
axis background/.style={fill=white},
axis x line*=bottom,
axis y line*=left,
axis line style={draw=none},
xlabel style={font=\color{white!15!black},yshift=20pt}, xlabel={\relax},
ylabel style={font=\color{white!15!black},yshift=-40pt}, ylabel={\relax},
xmajorticks=false,
ytick={5},
yticklabels={$q_2$},
y tick label style={rotate=90},
legend style={
	at={(0.97,0.03)}, anchor=south east, 
	legend cell align=left, align=left, draw=white!15!black}
]

\addplot [draw=none, fill=white!15!black, forget plot] coordinates {(0, 9.3+9*0 ) (0, 0.7+9*0 ) (0.2, 0.7+9*0 ) (0.2, 9.3+9*0 )};

\addplot[only marks, mark=square*, mark options={}, mark size=\LNmksz, draw=white!90!black, fill=white!90!black] table[row sep=crcr]{%
x	y\\
1	1\\
1	2\\
1	3\\
1	4\\
1	5\\
1	6\\
1	7\\
1	8\\
1	9\\
2	1\\
2	2\\
2	3\\
2	4\\
2	5\\
2	6\\
2	7\\
2	8\\
2	9\\
3	1\\
3	2\\
3	3\\
3	4\\
3	5\\
3	6\\
3	7\\
3	8\\
3	9\\
4	1\\
4	2\\
4	3\\
4	4\\
4	5\\
4	6\\
4	7\\
4	8\\
4	9\\
5	1\\
5	2\\
5	3\\
5	4\\
5	5\\
5	6\\
5	7\\
5	8\\
5	9\\
6	1\\
6	2\\
6	3\\
6	4\\
6	5\\
6	6\\
6	7\\
6	8\\
6	9\\
7	1\\
7	2\\
7	3\\
7	4\\
7	5\\
7	6\\
7	7\\
7	8\\
7	9\\
8	1\\
8	2\\
8	3\\
8	4\\
8	5\\
8	6\\
8	7\\
8	8\\
8	9\\
9	1\\
9	2\\
9	3\\
9	4\\
9	5\\
9	6\\
9	7\\
9	8\\
9	9\\
10	1\\
10	2\\
10	3\\
10	4\\
10	5\\
10	6\\
10	7\\
10	8\\
10	9\\
11	1\\
11	2\\
11	3\\
11	4\\
11	5\\
11	6\\
11	7\\
11	8\\
11	9\\
12	1\\
12	2\\
12	3\\
12	4\\
12	5\\
12	6\\
12	7\\
12	8\\
12	9\\
13	1\\
13	2\\
13	3\\
13	4\\
13	5\\
13	6\\
13	7\\
13	8\\
13	9\\
14	1\\
14	2\\
14	3\\
14	4\\
14	5\\
14	6\\
14	7\\
14	8\\
14	9\\
15	1\\
15	2\\
15	3\\
15	4\\
15	5\\
15	6\\
15	7\\
15	8\\
15	9\\
16	1\\
16	2\\
16	3\\
16	4\\
16	5\\
16	6\\
16	7\\
16	8\\
16	9\\
17	1\\
17	2\\
17	3\\
17	4\\
17	5\\
17	6\\
17	7\\
17	8\\
17	9\\
18	1\\
18	2\\
18	3\\
18	4\\
18	5\\
18	6\\
18	7\\
18	8\\
18	9\\
19	1\\
19	2\\
19	3\\
19	4\\
19	5\\
19	6\\
19	7\\
19	8\\
19	9\\
20	1\\
20	2\\
20	3\\
20	4\\
20	5\\
20	6\\
20	7\\
20	8\\
20	9\\
21	1\\
21	2\\
21	3\\
21	4\\
21	5\\
21	6\\
21	7\\
21	8\\
21	9\\
22	1\\
22	2\\
22	3\\
22	4\\
22	5\\
22	6\\
22	7\\
22	8\\
22	9\\
23	1\\
23	2\\
23	3\\
23	4\\
23	5\\
23	6\\
23	7\\
23	8\\
23	9\\
24	1\\
24	2\\
24	3\\
24	4\\
24	5\\
24	6\\
24	7\\
24	8\\
24	9\\
25	1\\
25	2\\
25	3\\
25	4\\
25	5\\
25	6\\
25	7\\
25	8\\
25	9\\
26	1\\
26	2\\
26	3\\
26	4\\
26	5\\
26	6\\
26	7\\
26	8\\
26	9\\
27	1\\
27	2\\
27	3\\
27	4\\
27	5\\
27	6\\
27	7\\
27	8\\
27	9\\
28	1\\
28	2\\
28	3\\
28	4\\
28	5\\
28	6\\
28	7\\
28	8\\
28	9\\
29	1\\
29	2\\
29	3\\
29	4\\
29	5\\
29	6\\
29	7\\
29	8\\
29	9\\
30	1\\
30	2\\
30	3\\
30	4\\
30	5\\
30	6\\
30	7\\
30	8\\
30	9\\
31	1\\
31	2\\
31	3\\
31	4\\
31	5\\
31	6\\
31	7\\
31	8\\
31	9\\
32	1\\
32	2\\
32	3\\
32	4\\
32	5\\
32	6\\
32	7\\
32	8\\
32	9\\
33	1\\
33	2\\
33	3\\
33	4\\
33	5\\
33	6\\
33	7\\
33	8\\
33	9\\
};
\addlegendentry{data1}

\addplot[only marks, mark=square*, mark options={}, mark size=\LNmksz, draw=white, fill=white] table[row sep=crcr]{%
x	y\\
28	3\\
30	1\\
25	2\\
25	3\\
26	1\\
26	3\\
27	1\\
27	2\\
22	1\\
22	2\\
22	3\\
23	1\\
23	2\\
23	3\\
24	1\\
24	2\\
24	3\\
19	1\\
19	2\\
19	3\\
20	1\\
20	2\\
20	3\\
21	1\\
21	2\\
21	3\\
16	1\\
16	2\\
16	3\\
17	1\\
17	2\\
17	3\\
18	1\\
18	2\\
18	3\\
13	1\\
13	2\\
13	3\\
14	1\\
14	2\\
14	3\\
15	1\\
15	2\\
15	3\\
10	1\\
10	2\\
10	3\\
11	1\\
11	2\\
11	3\\
12	1\\
12	2\\
7	1\\
7	2\\
7	3\\
8	1\\
8	2\\
9	1\\
4	1\\
4	2\\
5	1\\
1	1\\
28	6\\
30	4\\
25	6\\
27	4\\
22	6\\
24	4\\
19	6\\
21	4\\
16	6\\
18	4\\
13	6\\
15	4\\
10	6\\
12	4\\
7	6\\
9	4\\
};
\addlegendentry{data2}

\addplot[only marks, mark=+, mark options={}, mark size=\LNmksz, draw=RoyalBlue, fill=white] table[row sep=crcr]{%
x	y\\
28	3\\
30	1\\
25	2\\
25	3\\
26	1\\
26	3\\
27	1\\
27	2\\
22	1\\
22	2\\
22	3\\
23	1\\
23	2\\
23	3\\
24	1\\
24	2\\
24	3\\
19	1\\
19	2\\
19	3\\
20	1\\
20	2\\
20	3\\
21	1\\
21	2\\
21	3\\
16	1\\
16	2\\
16	3\\
17	1\\
17	2\\
17	3\\
18	1\\
18	2\\
18	3\\
13	1\\
13	2\\
13	3\\
14	1\\
14	2\\
14	3\\
15	1\\
15	2\\
15	3\\
10	1\\
10	2\\
10	3\\
11	1\\
11	2\\
11	3\\
12	1\\
12	2\\
7	1\\
7	2\\
7	3\\
8	1\\
8	2\\
9	1\\
4	1\\
4	2\\
5	1\\
1	1\\
28	6\\
30	4\\
25	6\\
27	4\\
22	6\\
24	4\\
19	6\\
21	4\\
16	6\\
18	4\\
13	6\\
15	4\\
10	6\\
12	4\\
7	6\\
9	4\\
};
\addlegendentry{data3}

\addplot[only marks, mark=square*, mark options={}, mark size=\LNmksz, draw=white, fill=white] table[row sep=crcr]{%
x	y\\
31	1\\
31	2\\
31	3\\
32	1\\
32	2\\
32	3\\
33	1\\
33	2\\
33	3\\
28	1\\
28	2\\
29	1\\
29	2\\
29	3\\
30	2\\
30	3\\
25	1\\
26	2\\
27	3\\
28	4\\
28	5\\
29	4\\
29	5\\
29	6\\
30	5\\
30	6\\
25	4\\
25	5\\
26	4\\
26	5\\
26	6\\
27	5\\
27	6\\
22	4\\
22	5\\
23	4\\
23	5\\
23	6\\
24	5\\
24	6\\
19	4\\
19	5\\
20	4\\
20	5\\
20	6\\
21	5\\
21	6\\
16	4\\
16	5\\
17	4\\
17	5\\
17	6\\
18	5\\
18	6\\
13	4\\
13	5\\
14	4\\
14	5\\
14	6\\
15	5\\
15	6\\
10	4\\
10	5\\
11	4\\
11	5\\
11	6\\
12	5\\
7	4\\
7	5\\
8	4\\
8	5\\
4	4\\
4	5\\
5	4\\
1	4\\
31	7\\
31	8\\
31	9\\
32	7\\
32	9\\
33	7\\
33	8\\
33	9\\
28	7\\
28	8\\
28	9\\
29	7\\
29	8\\
29	9\\
30	7\\
30	8\\
30	9\\
25	7\\
25	8\\
25	9\\
26	7\\
26	8\\
26	9\\
27	7\\
27	8\\
27	9\\
22	7\\
22	8\\
22	9\\
23	7\\
23	8\\
23	9\\
24	7\\
24	8\\
24	9\\
19	7\\
19	8\\
19	9\\
20	7\\
20	8\\
20	9\\
21	7\\
21	8\\
21	9\\
16	7\\
16	8\\
16	9\\
17	7\\
17	8\\
17	9\\
18	7\\
18	8\\
18	9\\
13	7\\
13	8\\
13	9\\
14	7\\
14	8\\
14	9\\
15	7\\
15	8\\
15	9\\
10	7\\
10	8\\
10	9\\
11	7\\
11	8\\
11	9\\
12	7\\
12	8\\
7	7\\
7	8\\
7	9\\
8	7\\
8	8\\
9	7\\
4	7\\
4	8\\
5	7\\
1	7\\
};
\addlegendentry{data4}

\addplot[only marks, mark=triangle, mark options={rotate=90}, mark size=\LNmksz, draw=RedOrange, fill=white] table[row sep=crcr]{%
x	y\\
31	1\\
31	2\\
31	3\\
32	1\\
32	2\\
32	3\\
33	1\\
33	2\\
33	3\\
28	1\\
28	2\\
29	1\\
29	2\\
29	3\\
30	2\\
30	3\\
25	1\\
26	2\\
27	3\\
28	4\\
28	5\\
29	4\\
29	5\\
29	6\\
30	5\\
30	6\\
25	4\\
25	5\\
26	4\\
26	5\\
26	6\\
27	5\\
27	6\\
22	4\\
22	5\\
23	4\\
23	5\\
23	6\\
24	5\\
24	6\\
19	4\\
19	5\\
20	4\\
20	5\\
20	6\\
21	5\\
21	6\\
16	4\\
16	5\\
17	4\\
17	5\\
17	6\\
18	5\\
18	6\\
13	4\\
13	5\\
14	4\\
14	5\\
14	6\\
15	5\\
15	6\\
10	4\\
10	5\\
11	4\\
11	5\\
11	6\\
12	5\\
7	4\\
7	5\\
8	4\\
8	5\\
4	4\\
4	5\\
5	4\\
1	4\\
31	7\\
31	8\\
31	9\\
32	7\\
32	9\\
33	7\\
33	8\\
33	9\\
28	7\\
28	8\\
28	9\\
29	7\\
29	8\\
29	9\\
30	7\\
30	8\\
30	9\\
25	7\\
25	8\\
25	9\\
26	7\\
26	8\\
26	9\\
27	7\\
27	8\\
27	9\\
22	7\\
22	8\\
22	9\\
23	7\\
23	8\\
23	9\\
24	7\\
24	8\\
24	9\\
19	7\\
19	8\\
19	9\\
20	7\\
20	8\\
20	9\\
21	7\\
21	8\\
21	9\\
16	7\\
16	8\\
16	9\\
17	7\\
17	8\\
17	9\\
18	7\\
18	8\\
18	9\\
13	7\\
13	8\\
13	9\\
14	7\\
14	8\\
14	9\\
15	7\\
15	8\\
15	9\\
10	7\\
10	8\\
10	9\\
11	7\\
11	8\\
11	9\\
12	7\\
12	8\\
7	7\\
7	8\\
7	9\\
8	7\\
8	8\\
9	7\\
4	7\\
4	8\\
5	7\\
1	7\\
};
\addlegendentry{data5}

\addplot[only marks, mark=square, mark options={}, mark size=\LNmksz, draw=RoyalBlue] table[row sep=crcr]{%
x	y\\
28	3\\
30	1\\
25	3\\
27	1\\
22	3\\
24	1\\
19	3\\
21	1\\
16	3\\
18	1\\
13	3\\
15	1\\
10	3\\
12	1\\
7	3\\
9	1\\
28	6\\
30	4\\
25	6\\
27	4\\
22	6\\
24	4\\
19	6\\
21	4\\
16	6\\
18	4\\
13	6\\
15	4\\
10	6\\
12	4\\
7	6\\
9	4\\
};
\addlegendentry{data6}

\legend{}; 
\end{axis}
\begin{axis}[%
width=4.521in,
height=1.33in,
at={(0.758in,1.599in)},
scale only axis,
xmin=0,
xmax=34,
ymin=0,
ymax=10,
axis x line*=bottom,
axis y line*=right,
axis line style={draw=none},
xlabel style={font=\color{white!15!black},yshift=20pt}, xlabel={\relax},
ylabel style={font=\color{white!15!black},yshift=40pt}, ylabel={\relax},
xmajorticks=false,
ytick={2,5,8},
yticklabels={$\var{g}_0$, $\var{g}_1$, $\var{g}_2$},
y tick label style={rotate=90},
legend style={
	at={(0.97,0.03)}, anchor=south east, 
	legend cell align=left, align=left, draw=white!15!black}
]

\addplot [draw=none, fill=white!15!black, forget plot] coordinates {(34-0.2+0, 3.3+3*0 ) (34-0.2+0, 0.7+3*0 ) (34-0.2+0.2, 0.7+3*0 ) (34-0.2+0.2, 3.3+3*0 )};
\addplot [draw=none, fill=white!15!black, forget plot] coordinates {(34-0.2+0, 3.3+3*1 ) (34-0.2+0, 0.7+3*1 ) (34-0.2+0.2, 0.7+3*1 ) (34-0.2+0.2, 3.3+3*1 )};
\addplot [draw=none, fill=white!15!black, forget plot] coordinates {(34-0.2+0, 3.3+3*2 ) (34-0.2+0, 0.7+3*2 ) (34-0.2+0.2, 0.7+3*2 ) (34-0.2+0.2, 3.3+3*2 )};

\end{axis}
\end{tikzpicture}%
}\vspace{-1pt}\\
	\resizebox{0.49\columnwidth}{!}{\input{tikz/TKZ_Q1.tex}}\hfill%
	\resizebox{0.49\columnwidth}{!}{\input{tikz/TKZ_Q0.tex}}\vspace{-5pt}
	\caption{Partitioning the state-space of the running example using
	$q_0$, $q_1$, $q_2$, and $q_3$.
	\HL{For example, $\St{q_1, \var{g}_0, \var{h}_5, 2, 1} = \AWbd $ indicates that
	$s:\St{g,h,x,y}=\St{0,5,2,1} \in R_{q_1}^{q_3} $.}
	}
	\label{fig:partitioning}
\end{figure}

\begin{definition}[System-CAPTL Composition]\label{def:srs} 
Let $\MDP = \MDPdef $, 
 $\CAPTL = \CAPTLdef $,
 and $ \Str = \{ \Str_\St{q,s_0} \mid q \in Q \} $.
Their \emph{composition}
 is defined as the automaton
 $
	\MDP_\CAPTL^\Str =  
	(V, \ActW , \Pmatrix_v, \rightarrow^\prime , \allowbreak
		v_0, \AP,
		\bar{L} )
 $
 where $ V \subseteq S \times Q \times \Playerset $,
 and $\Playerset = \left\{ \Pone, \Ptwo \right\} $;
 $\ActW = \Act \cup W \cup \{ \tau \} $, 
 where $\tau $ is a stutter action%
 ;
 $ v_0 = \St{ s_0, q_0, \Ptwo }$;
 $ \bar{L} : V \rightarrow \Pset(\AP) $ such that
 $ \bar{L} (\St{s,q,\gamma}) = L(s) $;
 and the transition relation $ \rightarrow^\prime $
 is defined using the following compositional rules:\\
\resizebox{\columnwidth}{!}{%
$$
\inference[\Rule1]
	{ s \xrightarrow{a,p} s^\prime  \land \Str_\St{q, s_0}(s) \!=\! a }
	{\St{s , q, \Pone } \xrightarrow{a,p}^\prime \St{s^\prime, q, \Ptwo} } \;
\inference[\Rule2]
	{s \in R_q^{q} }
	{\St{s, q,\Ptwo} \xrightarrow{\tau}^\prime \St{s, q, \Pone}} \;
\inference[\Rule3]
	{s \in R_q^{q'} }
	{\St{s, q,\Ptwo} \xrightarrow{w_\St{q,q'}}^\prime \St{s, q', \Ptwo} }.
$$ %
}
\end{definition}

The rules in~\defref{def:srs} are interpreted as follows.
The state space $V$ is partitioned into $V_\Pone$ (where $\MDP$ actions are allowed)
 and $V_\Ptwo$ (where $\CAPTL$ actions are allowed), resembling a turn-based 2-player
 game.
\Rule1 ensures that, if $q$ is active in $s$, then only the transitions with 
 the optimal action $\Str_\St{q, s_0}(s)$ are allowed.
\Rule2 ensures that, if $s \in R_q^q$, the active
 objective remains unchanged.
If $s \in R_q^{q'}$, however,
 \Rule3 enforces switching the active objective to $q'$. 
The action $\tau$ is a stutter since $\forall v \xrightarrow{\tau} v' $,
 $\bar{L}(v) = \bar{L}(v')$.

\begin{lemma}[Induced DTMC]\label{lem:dtmc} 
$\MDP_\CAPTL^\Str$ constructed using \defref{def:srs} is
 a DTMC.
\end{lemma}

\lemref{lem:dtmc} dictates that the probability measure $ \Probability_\SRS $
 is well-defined.
We will now use the notion of \emph{stutter equivalence}~\cite{baier2008principles}
 to prove that $\SRS$ 
 is bisimilar to $\MDPxPRO $.
Basically, two paths $\pi_1 $ and $\pi_2 $ are stutter-equivalent,
 denoted by $ \pi_1 \triangleq \pi_2 $, if there exists
 a finite sequence $ A_0 \ldots A_n \in (\Pset(\AP))^+ $ such that
 $\Trace (\pi) , \Trace(\hpi) \in A_0^+ A_1^+ \ldots A_n^+ $,
 \HL{where $A^{+} \Eq \left\lbrace A, AA, \ldots \right\rbrace $ is the set of finite,
 non-empty repetitions.}

\begin{theorem}[Stutter-Equivalence]\label{thm:equivalence} 
Let $\MDP$, $\CAPTL$,
 and $\Protocol \in \Protocolset$ be such that $\MDP, \Protocol \models \CAPTL $.
For every $ \pi \in \Fpath_{\MDPxPRO} $ there exists 
 $\hpi \in \Fpath_{\SRS} $ such that $ \pi \triangleq \hpi $ and
 $ \Probability_{\MDP^\Protocol} (\pi) = \Probability_{\SRS} (\hpi)$.
For every $ \hpi \in \Fpath_{\SRS} $, where $\Last(\hpi) \in V_\Ptwo $,
 there exists 
 $\hpi \in \Fpath_{\MDPxPRO} $ such that $ \hpi \triangleq \pi $ and
 $ \Probability_{\SRS} (\hpi) = \Probability_{\MDPxPRO} (\pi)$.
\end{theorem}

\begin{proofsketch}
We show that for every execution fragment $\varrho_1 = \St{s,q} \xra{a,p} \St{s,q'} $ 
 there exists
 $ \hat{\varrho}_1 = \St{s,q,\Ptwo} \xra{\tau} \St{s,q,\Pone} \xra{a,p} \St{s',q,\Ptwo} $.
Moreover, for every $\varrho_2 = \St{s,q} \xra{w} \St{s',q} $ there exists
 $ \hat{\varrho}_2 =  \St{s,q, \Ptwo} \xra{w} \St{s,q', \Ptwo} $. 
Using induction, we show that for every arbitrary execution $\varrho $ there exists
 $\hat{\varrho} $ such that $ \varrho \triangleq \hat{\varrho} $,
 where 
\begin{align*}
\begin{array}{llllllllll}
	\Trace(\varrho) &= 
		&(A_0+A_0A_0) &(A_1 + A_1 A_1) &\ldots &(A_n + A_n A_n) &\in (\Pset(AP))^+ \\
	\Trace(\hat{\varrho}) &=
		&(A_0 A_0 ) &(A_1 A_1) &\ldots &(A_n A_n) &\in (\Pset(AP))^+
\end{array}
\end{align*} 
 and $ \Probability(\varrho) = \Probability(\hat{\varrho})$.
Similarly, the other direction can be shown for every $\Last(\hat{\varrho})$
 that ends with $\Last(\hat{\varrho}) \in V_\Ptwo $.
\end{proofsketch}

We use both \lemref{lem:lsd} and \thmref{thm:equivalence} 
 to devise the protocol synthesis procedure summarized in \algref{alg:captl}.
In the first part (lines \ref{alg:captl:p1:start}--\ref{alg:captl:p1:end}),
 the procedure starts by 
 synthesizing a local strategy $\Str_\St{q_0,s_0}$ 
 and obtaining the associated LFP vector $\vec{x}_{q_0} \in \left[0,1 \right]^{R} $.
Next, $R$ is partitioned using $\vec{x}_{q_0}$ to obtain $(R_{q_0}^{q})_{q \in Q}$.
If $R_{q_0}^{q} \neq \emptyset $ for some $q \neq q_0$, the same procedure
 is repeated for $q$ to obtain 
 $\St{q,s_0}$, $\vec{x}_{q}$ and $(R_{q}^{q'})_{q' \in Q}$.
In the second part (lines \ref{alg:captl:p2:start}--\ref{alg:captl:p2:end}),
 three modules are constructed based on~\defref{def:srs}. 
The resulting parallel composition constitutes $\SRS$, 
 which mimics a stochastic 2-player game
 between $\MDPx$ (player $\Pone$) and $\CAPTLx$ (player $\Ptwo$),
 where the players' choices are already resolved by $\Strx$.
Finally, $\Protocol$ is populated by a query that checks for the CAPTL 
 satisfaction condition (line~\ref{alg:captl:last}), \ie
 a state $ \St{s,q_i, \gamma}$ is reached where
 $q_i = \Pquery_{\max} [\Eventually\Always B_i]$ is active, and $\Always B_i $ holds.
Notice that, based on the results from \lemref{lem:lsd},
 \algref{alg:captl} synthesizes a local strategy 
 at most once for every $q\in Q$,
 compared to \algref{alg:pctl} where synthesis is performed at every
 reachable state.

\begin{theorem}\label{thm:captl}
\algref{alg:captl} terminates; and returns $\Protocol, c$ iff 
 $\MDP,\Protocol \models_c \CAPTL $.
\end{theorem}

\begin{algorithm}[!tb]
\caption{Synthesis Procedure for P-CAPTL} \label{alg:captl}
	\KwIn{$\MDP = \MDPdef $, $\CAPTL = \CAPTLdef $}
	\KwResult{$\Protocol, c$ such that $\MDP, \Protocol \models_{c} \CAPTL$}
	
	\lForEach(\tcp*[f]{Initialize}){$ (q,q') \in Q \times Q $}{\label{alg:captl:p1:start}%
		$ R_q^{q'} \assign \emptyset $ 
	}
	$\Protocol \assign \emptyset $, \, 
	$\hat{Q} \assign  \left\{ q_0 \right\} $, \, 
	$\bar{Q} \assign  \emptyset $, \, 
	$ R \assign \Reachset(\MDP,s_0) $ \; 
	\While(\tcp*[f]{Partition $ R $}){%
		$\hat{Q} \neq \emptyset $ \label{alg:optimized:while}}{%
		Let $ q \in \hat{Q}$, \, 
		$\hat{Q} \assign \hat{Q} \setminus \left\{ q \right\} $, \,
		$\bar{Q} \assign \bar{Q} \cup \{ q \} $,
		$ R_q^q \assign R $ \;

		$ \Str_\St{q, s_0} \assign \Synth \left( \MDP; s_0, q \right) $,\,
		$ \vec{x}_q \assign \ReachPset \left( \MDP, s_0 , \Str_\St{q, s_0} \right) $ \;

		\ForEach(){$ w \in W(q) $ where $q' = Q(q,w) $}{%
			$ R_q^{q'} \assign \left\{ s \mid \vec{x}_q[s] \in J_w \right\} $,\,
			$ R_q^{q} \assign R_q^{q} \setminus R_q^{q'} $ \;
			\lIf(){$ R_q^{q'} \neq \emptyset \land q' \not\in \bar{Q} $ }{%
				$\hat{Q} \assign \hat{Q} \cup \left\{ q' \right\} $ }
		\RemoveAlgExtraSpace}
	\RemoveAlgExtraSpace} \label{alg:captl:p1:end}
	
	\Construct(\tcp*[f]{Construct $ \SRS $}){$\MDPx $ module}{\label{alg:captl:p2:start}%
		\lForEach(){$ [a]\, s \ra p_i : (s'_i) $}{%
			add $ [a]\, s \land \Pone \ra p_i : (s'_i) \land (\Ptwo) $}
	\RemoveAlgExtraSpace}
	\Construct{$\CAPTLx$ module}{%
		\lForEach(){$q \in \bar{Q} $}{%
			add $ [\tau]\, \qact = q \land \Ptwo \land L(R_q^{q};s) \ra
						(\qact=q) \land (\Pone) $}\RemoveAlgExtraSpace
		\lForEach(){$ q \xra{w} q'$}{%
			add $ [w]\, \qact\eqs q \land \Ptwo \land L(R_q^{q'};s) \ra 
						(\qact=q') \land (\Ptwo) $}
	\RemoveAlgExtraSpace}
	\Construct{$\Strx $ module}{%
		\lForEach(){$ \Str_\St{q,s_0} \neq \emptyset $ and $s \in R$}{%
			add $ [\Str_\St{q,s_0}(s)]\, \qact = q \land s \ra \top $}
	\RemoveAlgExtraSpace}
	
	$ \SRS \assign \MDPx \parallel \CAPTLx \parallel \Strx$%
	\label{alg:captl:p2:end} \; \RemoveAlgExtraSpace
	$ (\Protocol , c) \assign \Synth \left( \SRS,\, \St{q_0,s_0,\Ptwo},\, 
		\Pquery [ \bigvee_{q_i\in Q} \Eventually\Always (\qact=q_i ) \land B_i ] \right) $%
	\label{alg:captl:last}
\end{algorithm}

\begin{wrapfigure}[17]{r}{0.48\columnwidth} 
	\centering
	\vspace{-6ex} 
	\setlength{\LNmksz}{3.4pt}
	\pgfplotsset{every tick label/.append style={font=\LARGE}}
	\pgfplotsset{every legend/.append style={font=\LARGE}}
	\resizebox{0.46\columnwidth}{!}{\input{tikz/TKZ_Str.tex}}
	\vspace{-1ex}
	\caption{The protocol synthesized based on the CAPTL
	 requirement in~\figref{fig:ex:normal_form}, where $ R_V = \Reach(\SRS, v_0)$.}
	\label{fig:ex:synthesis}
\end{wrapfigure}

\refstepcounter{example}\noindent\textit{Example \theexample ~(Protocol Synthesis).}%
\label{ex:synthesis}
For the CAPTL requirement in \exref{ex:normal_form} (see~\figref{fig:ex:normal_form}), 
 \figref{fig:ex:synthesis}
 shows a visual representation of the protocol synthesized using~\algref{alg:captl}, 
 where blue markers indicate
 actions in $\Act$, and red markers indicate actions in $W$.
While pursuing $q_0$, the robot can achieve the task by moving 
 $\ANorth (\ANorthSym)$, $\ANorth (\ANorthSym)$,
 $\AEast (\AEastSym)$ 
 if no obstacles are encountered, or if obstacles are encountered only once
 while moving $\AEast (\AEastSym)$.
Switching from $q_0$ to $q_1$ via $w_{01} (\AWab)$ occurs in one state $(0,7,1,2)$; while
 switching from $q_0$ to $q_2$ via $w_{02} (\AWac)$ occurs in four states
 $(0,8,1,1)$, $(0,4,3,1)$, $(2,7,2,3)$ and $(0,4,1,3)$.

\section{Experimental Evaluation}
\label{sec:case}

We demonstrate the use of CAPTL for protocol synthesis and analysis on
 two case studies.
The first extends the robot task planning problem \HL{introduced}
 in~\secref{sec:problem}.
The second considers the problem of synthesizing an error-resilient
 scheduler for digital microfluidic biochips.
To this end, we implemented \algref{alg:captl} in \MATLAB{} on top of
 a modified version of \PRISMG{}~\cite{kwiatkowska2018prism} (v4.4),
 where $\ReachPset $ functionality was added.
The experiments presented in this section were run on an Intel Core i7 2.6GHz CPU
 with 16GB RAM.

\subsubsection{Robotic Task Planner.}

\tabref{tab:results} summarizes the
 performance results for running~\algref{alg:captl} on various sizes of the running
 example.
Notice that the number of choices in $\MDP_{\CAPTL}^{\Str}$ always 
 matches the number of states, which agrees with the results from~\lemref{lem:dtmc}.
In the three models, $q_0$ is always active in $s_0$, and thus is always
 verified.
As the grid size grows larger, the probability of reaching the goal --- and hence 
 satisfying $q_0$ --- becomes lower, dropping below $0.85$ at the initial state in
 both $(6 \!\times\! 6)$ and $(9 \!\times\! 9)$.
As a result, $q_1$ is never active (and hence is never verified) in the second and third
 models.
We also notice that the total time required to run~\algref{alg:captl} does
 not necessarily grow as the size of the problem grows.
In fact, the total time required for $(6 \!\times\! 6)$ and $(9 \!\times\! 9)$ is lower than
 the one for $(3 \!\times\! 3)$.
This is primarily due to the fact that $q_1$ is never reached or verified in the second
 and third models \HL{as we described.}
When comparing the model size for $\MDP$ and $\MDP_{\CAPTL}^{\Str}$, we notice
 that $|\MDP_{\CAPTL}^{\Str}| < |\MDP|$, with the difference being in orders of
 magnitude for larger models.
However, the time required to construct $\MDP_{\CAPTL}^{\Str}$ is longer than
 the time required to construct $\MDP$.

\subsubsection{MEDA-Biochip Scheduler.}

We now consider synthesizing error-resilient scheduler
 for micro-electrode-dot-array (MEDA) digital microfluidic biochips, where we borrow
 examples from~\cite{elfar2017synthesis,li2016error}.
A biochip segment consists of a $\var{W} \!\times\! \var{H}$ 
 matrix of on-chip actuators and sensors to 
 manipulate microfluidic droplets,
 and is further partitioned into $3\times 3$ blocks.
Two reservoirs are used to dispense droplets \kw{A} and \kw{B}.
Various activation patterns can be applied to manipulate the droplets, including
 moving (moving droplets individually), 
 flushing (moving both droplets at the same time in the same direction) and 
 mixing (merging two droplets occupying the same block).
As the biochip degrades, the actuators become less reliable,
 and an actuation command may
 not result in the droplet moving as expected.
The probability of an error occurring is proportional to
 the total number of errors occurred in the same block.

\begin{figure}[t]
	\centering
	\scalebox{1}{\begin{tikzpicture}[smallautomaton]
	\def\DeltaX{1cm}
	\begin{scope}[shift={(0,0)},on grid]
		\node[initial,state] (s0) at (0,0) {0};
		\node[state,below=25pt of s0] (s1) {1};
		\node[state,below right=15pt and 50pt of s1] (s2) {2};
		\node[state,below left=15pt and 50pt of s2] (s3) {3};
		\node[state,below=30pt of s3] (s4) {4};
		\node[state,left=50pt of s4] (s5) {5};
		\node[state,right=50pt of s4] (s6) {6};
		\node[state,right=50pt of s0] (s7) {7};
		
		\node[pLabel,below left=13pt and 2pt of s5] {$\{ \kwm{mixed} \} $};
		\node[pLabel,below right=13pt and 8pt of s7] {$\{ \kwm{aborted} \} $};
		\node[pLabel,below right=13pt and 6pt of s6] {$\{ \kwm{salvaged} \} $};
		
		\path[->]
			(s0) edge[ ] 
						node[pAction,left] {$\kwm{dispense}$!}  (s1)
			(s1) edge[out=0,in=135] 
						node[pAction,above right,outer sep=-2pt,pos=.8] {$\kwm{mvA}[d]!$} (s2)
			(s2) edge[out=-135,in=0] 
						node[pAction,below right,outer sep=-2pt,pos=.2] {$\kwm{mvB}[d]!$} (s3)
			(s3) edge[ ] 
						node[pAction,right,pos=.3] {$\kwm{update}!$} (s4)
			(s4) edge[ ] 
						node[pGuard,above,pos=.5] {$\kwm{inBlock}$} 
						node[pAction,below,pos=.3]{$\kwm{mix}$} (s5)
			(s4) edge[out=135,in=-135,looseness=1.4] 
						node[pGuard,left,pos=.6] {$\lnot \kwm{abAbsent}$} 
						node[pAction,left,pos=.4] {$\kwm{repeat}$} (s1)		
			(s1) edge[ ] 
						node[pAction,right] {$\kwm{flush}!$} (s3)
			(s4) edge[ ] 
						node[pGuard,above] {$\kwm{abAbsent}$} 
						node[pAction,below,pos=.3] {$\kwm{exit}$} (s6)
			(s0) edge[ ] 
						node[pAction,pos=.4] {$\kwm{abort}$} (s7)
			(s1) edge[bend right=5] 
						node[pAction,above left,outer sep=-2pt,pos=.6] {$\kwm{abort}$} (s7)
			(s5) edge[out=150,in=-150,looseness=5] node[above,pos=.5] { } (s5)
			(s6) edge[out=30,in=-30,looseness=5] node[above,pos=.5] { } (s6)
			(s7) edge[out=30,in=-30,looseness=5] node[above,pos=.5] { } (s7)
			;
	\end{scope}
	\begin{scope}[shift={(5.5cm,0)},on grid]
		\node[initial,state] (s0) at (0,0) {0};
		\node[dot, below=20pt of s0]  (p0) {};		
		\node[state,below=50pt of s0] (s1) {1};
		\node[dot,below=35pt of s1]   (p1) {};
		\node[dot,right=50pt of s1] (p2) {};		
		\node[state,below right=35pt and 50pt of s1] (s2) {2};
		\node[state,below left=35pt and 50pt of s1]  (s3) {3};
		
		\path[-]
			(s0) edge[shorten >=0pt] 
						node[pAction,right,outer sep=-1pt,pos=.6] {$\kwm{dispense}?$} (p0)
			(s1) edge[shorten >=0pt] 
						node[pAction,left,outer sep=-2pt,pos=.4] {$\kwm{mvA}[d]?$} (p1)
			(s1) edge[shorten >=0pt] 
						node[pAction,above,outer sep=-2pt] {$\kwm{flush}?$} (p2)
			;
			
		\path[->]
			(p0) edge[ ] 
						node[pAssign,right,outer sep=-1pt,align=left,pos=.4] 
							{$\begin{array}{lllll}
								x &\!=\! x_0 &\!+\! &\epsilon_x \\
     						y &\!=\! y_0 &\!+\! &\epsilon_y
     						\end{array}$} (s1)
     	(p1) edge[ ] 
     				node[pDistr,above,outer sep=-1pt,pos=.35] {$p_1(e_{\ell})$}
     				node[pAssign,below,align=left] 
     					{$\begin{array}{lll}
     						x & \!+\!\!=\! 0 \\ 
     						y & \!+\!\!=\! 0 
     						\end{array}$}
     				node[inner sep=0mm,pos=.05] (p1s2) {} 
     				 (s2)
     	(p1) edge[ ] 
     				node[pDistr,above,outer sep=-1pt,pos=.45] {$1 \!-\! p_1(e_{\ell})$}
     				node[inner sep=0mm,pos=.05] (p1s3) {} 
     				node[pAssign,below,align=left] 
     					{$\begin{array}{lll}
     						x & \!+\!\!=\! \Delta_x(d) \\ 
     						y & \!+\!\!=\! \Delta_y(d) 
     						\end{array}$} (s3)
     	(s2) edge[bend left=10] 
     				node[pAction,above,outer sep=10pt,pos=.3] {$\kwm{update}?$}
     				node[pAssign,above,outer sep=3pt,pos=.3] {$e_{\ell} \Plusplus $} (s1)
     	(s3) edge[out=60,in=180,looseness=1.4] 
     				node[pAction,above,outer sep=0pt,pos=.7] {$\kwm{update}?$}
     			 	node[pGuard,above,outer sep=9pt,pos=.7] {$\lnot \kwm{atExit}$} (s1)
     	(s3) edge[out=90,in=-150,looseness=1.4] 
     				node[pAction,above,outer sep=8pt,pos=.6] {$\kwm{update}?$} 
     				node[pGuard,above,outer sep=17pt,pos=.6] {$\kwm{atExit}$} (s0)
     	(p2) edge[out=90,in=0,looseness=1.4] 
     				node[inner sep=0mm,pos=0,yshift=2pt] (p2s0) {} 
     				node[pDistr,right,outer sep=-1pt,align=left,pos=.1] {$1 \!-\! p_2(e_{\ell})$}
     				node[pAssign,above right,outer sep=-2pt,align=left] {$x=0 $\\ $y=0$} (s0)
     	(p2) edge[out=-90,in=90]
     				node[inner sep=0mm,pos=0,yshift=-2pt] (p2s2) {} 
     				node[pDistr,right,outer sep=-1pt,align=left,pos=.4] {$p_2(e_{\ell})$} (s2)
     	;
     
     \path[pArc] (p1s2) edge [out=-90,in=-90,looseness=1.5]  (p1s3) ;
     \path[pArc] (p2s0) edge [out=0,in=0,looseness=1.5]  (p2s2) ;
     
	\end{scope}
	\end{tikzpicture}}\RemoveSpaceAfterTikz
	\caption{The MEDA biochip scheduler model (left) and the droplet model (right).}
	\label{fig:meda}
\end{figure}
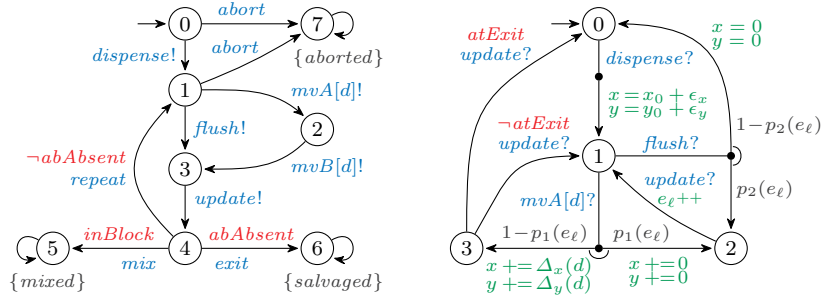

\figref{fig:meda} shows part of the segment scheduler (left) and the droplet (right)
 models.
In the initial \HL{state,} 
 the scheduler can dispense
 both droplets through the $\kwm{dispense}$ action, where the droplet location
 $(x,y)$ can probabilistically deviate from the dispenser location 
 $(x_0,y_0)$ with error $\epsilon$.
Subsequently, droplets can be individually manipulated via 
 $\kwm{mvA}[d]$ and $\kwm{mvB}[d]$ actions where $d$ is the direction, or together via \kwt{flush}.
The probability of successful manipulation (\ie $1-p(e_{\ell})$) depends on both 
 the number of errors occurred within the same block (\ie $e_{\ell}$) 
 and the activation pattern used.
The scheduler executes $\kwm{update}$ to sense droplet locations and register errors.

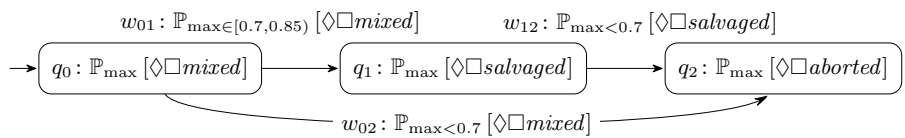
\begin{figure}[t]
	\centering
	\scalebox{0.95}{\begin{tikzpicture}[automaton]
	\def\DeltaX{1.1cm}
	\begin{scope}[shift={(0,0)}]
		\node[objective,initial] (q0) at (0,0) 
			{$q_0 \colon \Pquery_{\max} 
						\left[ \Eventually\Always \mathit{mixed} \right]$};
		\node[objective,right=\DeltaX of q0] (q1) 
			{$q_1 \colon \Pquery_{\max} 
						\left[ \Eventually\Always \mathit{salvaged} \right]$};
		\node[objective,right=\DeltaX of q1] (q2)  
			{$q_2 \colon \Pquery_{\max} 
						\left[ \Eventually\Always \mathit{aborted} \right]$};
		\path[->]
			(q0) edge[out=0,in=180] node[above,pos=.1,yshift=10pt]
				{$w_{01} \colon \Pquery_{\max \in \left[0.7,0.85 \right) } 
						\left[ \Eventually\Always \mathit{mixed} \right] $} (q1)
			(q1) edge[out=0,in=180] node[above,pos=.7,yshift=10pt]
				{$w_{12} \colon \Pquery_{\max <0.7} 
						\left[ \Eventually\Always \mathit{salvaged} \right] $} (q2)
			(q0) edge[out=-60,in=-120,looseness=0.2] node[pos=.5,fill=white]
				{$w_{02} \colon \Pquery_{\max <0.7} 
						\left[ \Eventually\Always \mathit{mixed} \right] $} (q2);
	\end{scope}
	\end{tikzpicture}}\RemoveSpaceAfterTikz
	\caption{P-CAPTL requirement for a MEDA-biochip segment scheduler.}
	\label{fig:meda_captl}
\end{figure}

The primary task of the scheduler is to perform a mixing operation within
 the given segment ($q_0$).
However, if the droplets are dispensed and (due to faulty blocks) the probability
 of a successful mixing operation is below $0.85$ ($w_{01}$), 
 salvaging the dispensed droplets
 by moving them to an adjacent segment is prioritized ($q_1$).
If the mixing probability drops below $0.7$ ($w_{02}$),
 or if the salvaging probability drops below $0.7$ ($w_{12}$), the scheduler is to
 abort the operation ($q_2$).
The aforementioned requirements are formalized using CAPTL as shown 
 in~\figref{fig:meda_captl}.
The set of objectives is $Q = \{ q_0, q_1, q_2 \}$, and the set of contexts is defined as
 $W = \{ w_{01}, w_{02}, w_{12} \}$.
The performance results for running~\algref{alg:captl} on three different segment sizes
 is reported in~\tabref{tab:results}. 

\begin{table*}[!tb] 
\caption{Protocol synthesis performance results for
	the robotic task planner (C1) and the MEDA-biochip scheduler (C2).
	(St.: states, Tr.: transitions, Ch.: choices).}
\label{tab:results}
\setlength{\tabcolsep}{3pt}%
\newcommand{\None}{\multicolumn{1}{c}{--}}%
\renewcommand{\arraystretch}{1.2}
\resizebox{\columnwidth}{!}{%
\begin{tabular}{r|rrr|rrr|rrrrrrrr}
	\hline	
	\multicolumn{1}{c}{Model} &
	\multicolumn{3}{|c}{$\MDP$ Size} &
	\multicolumn{3}{|c}{$\MDP_{\CAPTL}^{\Str}$ Size} &
	\multicolumn{8}{|c}{Construction/Synthesis Time (sec)} \\
	\hline
	\multicolumn{1}{c}{Size} &
	\multicolumn{1}{|c}{St.} &
	\multicolumn{1}{c}{Tr.} &
	\multicolumn{1}{c}{Ch.} &
	\multicolumn{1}{|c}{St.} &
	\multicolumn{1}{c}{Tr.} &
	\multicolumn{1}{c}{Ch.} &
	\multicolumn{1}{|c}{$\MDP$} &
	\multicolumn{1}{c}{$q_0$} &
	\multicolumn{1}{c}{$q_1$} &
	\multicolumn{1}{c}{$q_2$} &
	\multicolumn{1}{c}{$q_3$} &
	\multicolumn{1}{c}{$\MDP_{\CAPTL}^\Str$} &
	\multicolumn{1}{c}{$q_{\CAPTL}$} &
	\multicolumn{1}{c}{Total} \\
	\hline
	C1 $3 \!\times\! 3$
		& \num{233} & \num{1117} & \num{745} 
		& \num{142} & \num{163} & \num{142} 
		& {0.438} & {0.031} & {0.029} & {0.033} & {0.106} & {0.557} & {0.052} 
		& {25.5}  \\ 
	$6 \!\times\! 6$
		& \num{595} & \num{2692} & \num{1874} 
		& \num{159} & \num{190} & \num{159} 
		& {0.495} & {0.041} & \None & {0.083} & {0.260} & {0.662} & {0.112} 
		& {24.2}  \\
	$9 \!\times\! 9$
		& \num{733} & \num{3242} & \num{2278} 
		& \num{96} & \num{116} & \num{96} 
		& {0.508} & {0.037} & \None & {0.059} & {0.313} & {0.691} & {0.083} 
		& {21.9}  \\
\hline
	C2 $8 \!\times\! 5$
		& \num{2851} & \num{8269} & \num{5678} 
		& \num{2576} & \num{2929} & \num{2576} 
		& {1.308} & {2.348} & {0.433} & {3.122} & \None & {17.95} & {3.585} 
		& {60.53}  \\
	$11 \!\times\! 5$
		& \num{8498} & \num{25502} & \num{17214} 
		& \num{4167} & \num{4673} & \num{4167} 
		& {2.013} & {7.212} & {1.577} & {9.928} & \None & {79.77} & {5.84} 
		& {149.6}  \\
	$11 \!\times\! 8$
		& \num{15290} & \num{47602} & \num{31316} 
		& \num{3223} & \num{3653} & \num{3223} 
		& {2.065} & {12.36} & {2.536} & {18.61} & \None & {109.2} & {4.498} 
		& {218.5}  \\
	$14 \!\times\! 8$
		& \num{61489} & \num{201469} & \num{130718} 
		& \num{1016} & \num{1339} & \num{1016} 
		& {4.545} & {48.07} & {10.67} & {68.40} & \None & {289.9} & {1.289} 
		& {450.4}  \\
	\hline
\end{tabular}%
}%
\end{table*}

\section{Conclusion}
\label{sec:conclusion}

In this paper we have introduced context-aware probabilistic temporal logic (CAPTL).
The logic provides \HL{intuitive} means to formalize requirements that comprises a number
 of objectives with an underlying priority structure.
CAPTL allows for defining context (\ie probabilistic conditions) as the basis for
 switching between two different objectives.
We have presented CAPTL syntax and semantics for Markov-decision processes (MDPs).
We have also investigated the CAPTL synthesis problem, both from PCTL and CAPTL-based
 approaches, where we have shown that the latter provides significant performance
 improvements.
To demonstrate our work, we have presented two case studies.

As this work has primarily considered CAPTL semantics for MDPs, further investigation
 is required to generalize the results for stochastic multi-player games.
Another research direction involves expanding the results to include PCTL 
 fragments beyond persistence objectives, such as safety, bounded reachability and 
 reward-based objectives.

\vspace{-0.5ex}
\subsubsection*{Acknowledgments.}
	This work was supported in part by the NSF CNS-1652544 and ECCS-1914796, 
	ONR N00014-20-1-2745 and N00014-17-1-2504,
	as well as AFOSR FA9550-19-1-0169 awards.

{\linespread{0.88}
\bibliographystyle{splncs04}
\bibliography{Biblio_Elfar}
}
\begin{extended}
\newpage\appendix
\section{Proofs}
\label{app:proofs}

{\renewcommand{\theproposition}{\ref{pro:pctl}}
\begin{proposition}
\algref{alg:pctl} terminates; and returns $\Protocol, c$ iff 
 $\MDP,\Protocol \models_c \CAPTL $.
\end{proposition}
\addtocounter{proposition}{-1}}
\begin{proof}

We break the proof into two parts: termination and correctness.

\paragraph{Termination.}
We first note that $|S|, |Q|, |W| < \infty $ by definition.
\begin{itemize}
\item The \KwSty{foreach} loop (line~\ref{alg:pctl:foreach}) terminates either by
 exhausting all $w \in W(q) $ (which is finite), or by breaking whenever
 $ \VRepeat = \top$.
The loop can only run indefinitely if $|W(q)| = \infty$.
\item For the inner \KwSty{while} loop (line~\ref{alg:pctl:secondwhile}),
	the only way to remain indefinitely in that loop is for $\VRepeat = \top$ to 
	always hold, which is only set whenever $\MDP,s \models w$.
	However, whenever $\MDP,s \models w$ holds, $q$ is updated (line~\ref{alg:pctl:qupdate}).
	Since $Q$ is finite and $\CAPTL$ is acyclic, recursion over $Q$ ends in a finite 
	number of loops, ending with a $q \in Q$ where $W(q) = \emptyset$.
	Hence, $\MDP,s \models w$ cannot hold indefinitely.
\item For the outermost \KwSty{while} loop (line~\ref{alg:pctl:firstwhile}),
	line~\ref{alg:pctl:sremove} dictates that the set $S_q$ shrinks by one state $s$
	each and every loop, which is also added to $\bar{S}_q$.
	Hence, for $S_q \neq \emptyset $ to hold indefinitely for some $q\in Q$,
	$ \Post \left(\MDP, s,\Str(s) \right) \setminus \bar{S}_q \neq \emptyset $
	must always hold (line~\ref{alg:pctl:post}).
	However, this mandates that $\bar{S}_q$ can grow indefinitely.
	Since $|S| < \infty $ by definition, $ \bar{S}_q$ cannot grow indefinitely, and the
	loop eventually terminate in a finite number of iterations.	
\end{itemize} 

\paragraph{Correctness.}
Initially, $ s = s_0$ and $q = q_0$.
We identify the following cases:
\begin{itemize}
\item (a) Case $ W(q) = \emptyset $.
	Then $ \VRepeat = \bot $, $\Str$ is synthesized such that $\MDP^{\Str}_s \models q $,
	and $(s,q,\Str(s))$ is added to $\Protocol$.
\item (b) Case $ \forall w \in W(q) $, $ \MDP,s \not\models w $.
	Then $ \VRepeat = \bot $, $\Str$ is synthesized such that $\MDP^{\Str}_s \models q $,
	and $(s,q,\Str(s))$ is added to $\Protocol$.
\item (c) Case $\exists w \in W(q) $, $\MDP, s \models w $.
	Then from~\defref{def:normal_form} we conclude
	that $\forall \bar{w} \in W(q) \setminus \{ w \} $ it holds that 
	$\MDP, s \not\models \bar{w} $.
	Consequently, $ \VRepeat = \top $,
	$ (s,q,w) $ is added to $\Protocol$, and $q$ is updated.
	Since $\CAPTL$ is finite and acyclic, the loop eventually halts with condition (a) or (b)
	becoming true.
\end{itemize}%

\end{proof}

{\renewcommand{\thelemma}{\ref{lem:lsd}}
\begin{lemma}[Local Strategy Dominance]
Let $ \MDP = \MDPdef $ and $q=\Pquery_{\max} [\Eventually\Always B]$.
For all $ s \in \Reach(\MDP,s_0) $, 
 $\Str_\St{q,s} = \restr{\Str_\St{q,s_0} }{\Reach(\MDP,s)} $.
 
\end{lemma}
\addtocounter{lemma}{-1}}
\begin{proof} 

In the first part of the proof, we establish the used notation.
In the second part, we show that if $s \in \Reach(\MDP,s_0)$, then 
 the domain of $\Str_{\St{q,s}} $ is subset of
 the domain of $\Str_{\St{q,s_0}}$.
In the last part, we show that 
 $ \Str_\St{q,s} = \restr{\Str_\St{q,s_0} }{\Reach(\MDP,s)} $

\paragraph{Notation.}
For a function $ f: A \rightarrow B$,
 we will use $\Domain(f)$ to denote the domain of $f$, and
 $ f(a) \defined \iff a \in A$, $ f(a) \undefined \iff a \not\in A$.
We will use $\Strset_s : S \nrightarrow \Act $
 to denote the set of all possible (pure memoryless) strategies from state $s \in S$.
We assume that for every $\Str \in \Strset_s$, $\Str(s')$ is defined for every
 $s' \in \Reach(\MDP,s)$.

\paragraph{Well-Definedness.}
Now, let us consider an arbitrary strategy $ \Str_{s_0} \in \Strset_{s_0} $.
Hence, $\Domain(\Str_{s_0}) = \Reach(\MDP, s_0) $.
Moreover, consider an arbitrary state $ s \in \Reach(\MDP,s_0)$, and an associated
 strategy $ \Str_s \in \Strset_s $.
In such case, $ \Domain(\Str_s) = \Reach(\MDP,s)$.
Assume we can find a state $ s' \in S $ such that 
 $\Str_s (s') \defined $ and $\Str_{s_0} (s') \undefined $.
The assumptions imply that
\begin{align}\label{eq:reach:a}
\inference{\Str_s(s') \defined}{s' \in \Reach(\MDP,s)},\;
\inference{\Str_{s_0} (s') \undefined}{s' \not\in \Reach(\MDP,s_0)} 
\end{align}
However,
\begin{align}\label{eq:reach:b}
\inference{s \in \Reach(\MDP,s_0)}{\Reach(\MDP,s) \subseteq \Reach(\MDP,s_0)}
\end{align}
Since the inferences from \eqref{eq:reach:a} and \eqref{eq:reach:b} contradict
 each other, we conclude that 
 $ \Str_s (s') \defined \implies \Str_{s_0} (s') \defined $.
That is, for every $s \in \Reach(\MDP,s_0)$,
 $\Domain(\Str_s) \subseteq \Domain(\Str_{s_0})$.

\paragraph{Equivalence.}
We now prove that $\Str_\St{q,s}(s') = \Str_\St{q,s_0}(s')$ for every $ s' \in \Reach(\MDP,s)$.
To this end, we first recall a well-established result on the existence of 
 memoryless strategies for probabilistic reachability requirements.
For $\MDP$ and $q=\Pquery_{\max} [\Eventually\Always B]$, let $ R = \Reach(\MDP,s_0)$.
The maximum probability of reaching the target set $B$ from a state $s \in R $
 can be formulated as
\begin{align*}
	p_{\max} (s, q) = \sup_{\Str \in \Strset} \Probability^{\Str}_{\MDP,s} 
	\left( \left\{ s \ldots t_0 t_1 \ldots \in \Paths_{\MDP,s}^\Str | \forall i\geq 0, t_i \in B \right\} \right).
\end{align*}
%
As the computation of $p_{\max} (s', q)$ at state $s' \in \Reach(\MDP,s_0)$ is independent of the path
 that lead to $s'$, the optimal action associated with $s'$ is also independent of such
 path.
Therefore, $\Str_\St{q,s}(s') = \Str_\St{q,s_0}(s')$.

\end{proof}

{\renewcommand{\thelemma}{\ref{lem:ssp}} 
\begin{lemma}[Partitioning]
Let $ \MDP = \MDPdef $, $\CAPTL = \CAPTLdef$, and $R = \Reach(\MDP,s_0)$. 
For every $q \in Q$,
 $ \bigcup_{q' \in Q} R_q^{q'} \allowbreak = R $; and
 $ R_q^{q'} \cap R_q^{q''} = \emptyset $ for every $ q' \neq q'' $.
\end{lemma}
\addtocounter{lemma}{-1}}
\begin{proof}


For every $q\in Q$, we identify two cases:
\begin{itemize}
\item Case $W(q) = \emptyset $.
	In this case, $R_q^{q'} = \emptyset $ for every $q^\prime \neq q$,
	and $R_q^q = R \setminus \cap_{q' \neq q} R_q^{q'} = R$.
	Therefore, the lemma holds.
\item Case $W(q) \neq \emptyset $.
	By definition, $s \in R_q^q$ implies that $\vec{x}_q[s] \not\in J' $ for every
	$ w_{\langle q,q' \rangle} = \Pquery_{\max \in J'} [\Eventually\Always B] \in W(q)$.
	Therefore, $ R_q^q \cap R_q^{q'} = \emptyset $ holds for every $q' \neq q$ (a).
	Next, let us assume that we find $s \in R$ such that 
	$ s \in R_q^{q'}$ and $ s \in R_q^{q''}$, where $q' \neq q'' $ and
	$w_{\langle q,q'' \rangle} = \Pquery_{\max \in J''} [\Eventually\Always B] \in W(q)$.
	This implies that $ J' \cap J'' \neq \emptyset $,
	which contradicts~\defref{def:normal_form}.
	Hence, such $s$ does not exist, and $R_q^{q'} \cup R_q^{q''} = \emptyset $ holds
	for every $q' \neq q'' $(b).
	From (a) and (b), we conclude that the lemma holds.
\end{itemize}
 
\end{proof}

{\renewcommand{\thelemma}{\ref{lem:dtmc}} 
\begin{lemma}[Induced DTMC]
$\MDP_\CAPTL^\Str$ constructed using \defref{def:srs} is
 a DTMC.
\end{lemma}
\addtocounter{lemma}{-1}}
\begin{proof}
Assume that there exists a state $ v \in V $ with at least two actions
 $ \{\alpha_1, \alpha_2 \} \in \ActW(v) $. 
We identify two disjoint subsets of $V$, namely,
 $ V_\Pone = \{ v \in V \mid v = (s,q,\Pone) \}  $ and
 $ V_\Ptwo = \{ v \in V \mid v = (s,q,\Ptwo) \} $.
In case $v = (s,q,\Pone) \in V_\Pone $, only transitions defined by R1 are allowed.
Since for any $q$ at most one action $\Str_{\langle q, s_0 \rangle}(s)$ is allowed,
 we conclude that $v \not\in V_\Pone$.
In case $v = (s,q,\Ptwo) \in V_\Ptwo $, only transitions defined by R2 and R3 are allowed.
Our assumption requires that $ s \in S_q^{q} $ and $ s \in S_q^{q^\prime} $.
Since $ S_q^{q} \cap S_q^{q^\prime} = \emptyset $ by definition, it contradicts with
 our assumption, hence $v \not\in V_\Ptwo$.
Consequently, $v \not\in V$, which contradicts our assumption.
We conclude that $ \MDP_\CAPTL^\Str $ has no nondeterministic choices.
\end{proof}

{\renewcommand{\thetheorem}{\ref{thm:equivalence}}
\begin{theorem}[Stutter-Equivalence]
Let $\MDP$, $\CAPTL$,
 and $\Protocol \in \Protocolset$ be such that $\MDP, \Protocol \models \CAPTL $.
For every $ \pi \in \Fpath_{\MDPxPRO} $ there exists 
 $\hpi \in \Fpath_{\SRS} $ such that $ \pi \triangleq \hpi $ and
 $ \Probability_{\MDP^\Protocol} (\pi) = \Probability_{\SRS} (\hpi)$.
For every $ \hpi \in \Fpath_{\SRS} $, where $\Last(\hpi) \in V_\Ptwo $,
 there exists 
 $\hpi \in \Fpath_{\MDPxPRO} $ such that $ \hpi \triangleq \pi $ and
 $ \Probability_{\SRS} (\hpi) = \Probability_{\MDPxPRO} (\pi)$.
\end{theorem}
\addtocounter{theorem}{-1}}
\begin{proof}
The transitions of $ \MDP^\Protocol $ can be partitioned into two subsets
 $ \rightarrow_{\Act} $ and $ \rightarrow_{W} $ where the transitions take
 the forms 
 $ \St{s,q} \xrightarrow{a,p} \St{s',q} $ and
 $ \St{s,q} \xrightarrow{w} \St{s,q'} $, respectively.
Starting from $\St{s_0,q_0}$, let us assume that 
 $ \St{s_0,q_0} \xrightarrow{a,p} \langle s_1,q_0 \rangle $,
 which is based on $\Str_{\langle s_0, q_0 \rangle} (s_0)$.
Similarly, $ \SRS $ exhibits the execution fragment
 $$ \langle s_0, q_0, \Ptwo \rangle \xrightarrow{\tau} 
 \St{s_0, q_0, \Pone} \xrightarrow{\hat{a}, \hat{p} } 
 \langle \hat{s}_1, q_0, \Ptwo \rangle . $$
Since $ \hat{a} = \Str_{\langle s_0, q_0 \rangle } (s_0) $, 
 we conclude that $ \hat{a} = a $,
 and hence $ \hat{p} = p$ and $ \hat{s}_1 = s_1 $.
From \lemref{lem:lsd}, we know that 
 $\Str_{\langle s_0, q \rangle}(s) = \Str_{\St{s,q}}(s)$.
Hence, for every execution fragment $\varrho_\Act $ in $\MDP^\Protocol$, where no objective
 switching occurs, we can find an execution
 $\hat{\varrho}_\Act $ in $\SRS$ such that
\begin{align*}
\begin{array}{lllllllll}
	\varrho_\Act &= 
		&\St{s_0,q_0} &\xra{a_1,p_1} &\St{s_1,q_0} &\xra{a_2,p_2} 
		&\cdots &\St{s_i,q_0} \\
	\hat{\varrho}_\Act &=
		&\St{s_0,q_0,\Ptwo} \xra{\tau} \St{s_0,q_0,\Pone} &\xra{a_1,p_1} &\St{s_1,q_0,\Ptwo}
		&\xra{a_2,p_2} &\cdots &\St{s_i,q_0,\Ptwo},
\end{array}
\end{align*}
\begin{align*}
\begin{array}{llllllllll}
\Trace(\varrho_\Act ) &= 
	&\underbrace{L(s_0)}_{\mathrm{1-time}} 
	&\underbrace{L(s_1)}_{\mathrm{1-time}}
	&\cdots
	&\underbrace{L(s_i)}_{\mathrm{1-time}} \\
\Trace(\hat{\varrho}_\Act ) &= 
	&\underbrace{L(s_0) L(s_0)}_{\mathrm{2-times}} 
	&\underbrace{L(s_1) L(s_1)}_{\mathrm{2-times}}
	&\cdots
	&\underbrace{L(s_i) L(s_i)}_{\mathrm{2-times}} ,
\end{array}
\end{align*}
\begin{align*}
\begin{array}{llllllllll}
\Probability(\varrho_\Act ) &= &(p_0) &\cdot &(p_1) &\cdots &(p_i) \\
\Probability(\hat{\varrho}_\Act ) &= &(p_0) &\cdot &(p_1) &\cdots &(p_i) \\
\end{array}
\end{align*}
Therefore, $ \varrho_\Act \triangleq \hat{\varrho}_\Act $.
Now, consider an execution fragment that ends with switching the active objective.
In that case, for every execution fragment $\varrho $ we can find $\hat{\varrho}$
 such that
\begin{align*}
\begin{array}{lllllllll}
	\varrho_W &= 
		&\St{s,q} &\xra{w_{\St{q,q'}}} &\St{s,q'} \\
	\hat{\varrho}_W &=
		&\St{s,q,\Ptwo} &\xra{w_{\St{q,q'}}} &\St{s,q',\Ptwo},
\end{array}
\end{align*}
\begin{align*}
\begin{array}{lllllllll}
\Trace(\varrho_W ) &= 
	&\underbrace{L(s) L(s)}_{\mathrm{2-times}} \\
\Trace(\hat{\varrho}_W ) &= 
	&\underbrace{L(s) L(s)}_{\mathrm{2-times}},
\end{array}
\end{align*}
and
\begin{align*}
\begin{array}{llllllllll}
\Probability(\varrho_W ) &= 1 \\
\Probability(\hat{\varrho}_W ) &= 1 \\
\end{array}
\end{align*}
Therefore, $ \varrho_W \triangleq \hat{\varrho}_W $.
Using induction, we can show that for every arbitrary execution $\varrho $ there exists
 $\hat{\varrho} $ such that $ \varrho \triangleq \hat{\varrho} $,
 where 
\begin{align*}
\begin{array}{llllllllll}
	\Trace(\varrho) &= 
		&(A_0+A_0A_0) &(A_1 + A_1 A_1) &\ldots &(A_n + A_n A_n) &\in (\Pset(AP))^+ \\
	\Trace(\hat{\varrho}) &=
		&(A_0 A_0 ) &(A_1 A_1) &\ldots &(A_n A_n) &\in (\Pset(AP))^+
\end{array}
\end{align*} 
 and $ \Probability(\varrho) = \Probability(\hat{\varrho})$.
\end{proof}

{\renewcommand{\thetheorem}{\ref{thm:captl}}
\begin{theorem}
\algref{alg:captl} terminates; and returns $\Protocol, c$ iff 
 $\MDP,\Protocol \models_c \CAPTL $.
\end{theorem}
\addtocounter{theorem}{-1}}
\begin{proof}
We break the proof into two parts: termination and correctness.

\paragraph{Termination.}
We first note that $|S|, |Q|, |W| < \infty $ by definition.
\begin{itemize}
\item The \KwSty{foreach} loop (line~6) terminates by
 exhausting all $w \in W(q) $ (which is finite).
 The loop can only run indefinitely if $|W(q)| = \infty$.

\item For the \KwSty{while} loop (line~3),
	line~\ref{alg:pctl:sremove} dictates that the set $\hat{Q}$ shrinks by one element $q$
	each and every loop, which is also added to $\bar{Q}$.
	Moreover, line--8 dictates that an objective $q'$ is added to $\hat{Q}$ only if
	it is not in $\bar{Q}$.
	That is, every objective $q\in Q$ can be added at most once to $\bar{Q}$.
	Since $|Q| < \infty $ by definition, the condition $\hat{Q} = \emptyset $ is met in
	a finite number of iterations.

\item Since $ |\MDP| < \infty $, every \KwSty{construct} code blocks also terminates
 in a finite number of iterations.
 
\end{itemize} 

\paragraph{Correctness.}
From \thmref{thm:equivalence}, we know that the paths in $\MDPxPRO$ and $\SRS$ are 
 stutter equivalent and probabilistically bisimilar.
Hence, for every $q_i \in Q$, the two probability measures
\begin{align*}
\begin{array}{lll}
\Probability_\MDPxPRO &\left( \left\{ \pi \mid 
			\pi \models \Eventually  \St{ q_i,s} \land \vec{C}(q_i,s)=1 \right\} \right), \\
\Probability_\SRS &\left( \left\{ \hpi \mid 
			\hpi \models \Eventually\Always  (\qact = q_i) \land B_i \right\} \right)
\end{array}
\end{align*}
 are equivalent.
 Hence, the algorithm returns a correct answer.
\end{proof}

\end{extended}

\end{document}